\documentclass[fleqn,usenatbib]{mnras}

\usepackage{newtxtext,newtxmath}

\usepackage[T1]{fontenc}
\usepackage{ae,aecompl}

\usepackage{graphicx}	
\usepackage{amsmath}	
\usepackage{amssymb}	

\newcommand\un[1]{{\,\rm #1}}
\newcommand\E[1]{\times10^{#1}}
\newcommand\rs[1]{_\mathrm{#1}}

\newcommand\g{$\gamma$}

\title[Oblique post-adiabatic shocks in non-uniform media]{Post-adiabatic supernova remnants in an interstellar magnetic field: oblique shocks and non-uniform environment}

\author[O. Petruk, T. Kuzyo, S. Orlando,  M. Pohl, M. Miceli, F.Bocchino, V. Beshley, R. Brose]{O. Petruk$^{1,2}$, T. Kuzyo$^{1}$, S. Orlando$^{3}$, M. Pohl$^{4,5}$, M. Miceli$^{6,3}$, F. Bocchino$^{3}$,\newauthor V. Beshley$^{1}$, R. Brose$^{4,5}$\\
$^{1}$Institute for Applied Problems in Mechanics and Mathematics, Naukova 3-b, 79060 Lviv, Ukraine\\
$^{2}$Astronomical Observatory of the Jagiellonian University, Orla 171, 30-244 Krak\'ow, Poland\\
$^{3}$INAF - Osservatorio Astronomico, Piazza del Parlamento 1, 90134 Palermo, Italy\\
$^{4}$Universit\"at Potsdam, Institut f\"ur Physik \& Astronomie, Karl-Liebknecht-Strasse 24/25, 14476 Potsdam, Germany\\
$^{5}$ DESY, Platanenallee 6, 15738 Zeuthen, Germany\\
$^{6}$ Dipartimento di Fisica e Chimica, Universit\'a di Palermo, Piazza del Parlamento 1, 90134 Palermo, Italy}

\date{Accepted XXX. Received YYY; in original form ZZZ}

\pubyear{2018}

\begin{document}
\label{rad2:firstpage}
\pagerange{\pageref{rad2:firstpage}--\pageref{rad2:lastpage}}
\maketitle

\begin{abstract}
We present very-high-resolution 1D MHD simulations of the late-stage supernova remnants (SNR). In the post-adiabatic stage the magnetic field has an important and significant dynamical effect on the shock dynamics, the flow structure, and hence the acceleration and emission of cosmic rays. We find that the tangential component of the magnetic field provides pressure support that to a fair degree prevents the collapse of the radiative shell and thus limits the total compression ratio of the partially or fully radiative forward shock. A consequence is that the spectra of cosmic rays would not be as hard as in hydrodynamic simulations. We also investigated the effect on the flow profiles of the magnetic-field inclination and a large-scale gradient in the gas density and/or the magnetic field. A positive density gradient shortens the evolutionary stages whereas a shock obliquity lowers the shock compression. The compression of the tangential component of the magnetic field leads to its dominance in the downstream region of post-adiabatic shocks for a wide range of orientation of the upstream field, which may explain why one preferentially observes tangential radio polarization in old SNRs. As most cosmic rays are produced at late stages of SNR evolution, the post-adiabatic phase and the influence of the magnetic field during it are most important for modeling the cosmic-ray acceleration at old SNRs and the gamma-ray emission from late-stage SNRs interacting with clouds.
\end{abstract}



\section{Introduction}

Supernova remnants (SNRs) are the best laboratories for studying the physics of strong collision-less, magneto-hydrodynamic (MHD) shocks. These shocks effectively heat and compress the ambient medium, amplify magnetic field (MF), and accelerate electrons and ions up to very high energies, at which we refer to them as cosmic rays. 

The dynamics of accelerated charged particles may be studied by analyzing their photon emission. The radio emission of electrons in SNRs, including its polarization, has been observed for many decades \citep{2015A&ARv..23....3D}. Synchrotron emission of electrons at the highest energies gives rise to nonthermal X-rays which were first discovered in remnant of SN1006 in 1995 \citep{1995Natur.378..255K}, which implies the electrons emitting nonthermal X-rays should also reveal themselves in gamma rays on account of the inverse-Compton effect \citep{1996A&A...307L..57P}. In contrast, the only possibility to see emission due to accelerated protons is through high-energy $\gamma$-rays arising from proton-proton interactions. The detection of TeV-band gamma rays from the non-thermal SNR shell RX J1713.7-3946 \citep{2004Natur.432...75A} in 2004 opened a new era of gamma-ray astronomy, as, for the first time, detailed gamma-ray images of various astrophysical objects became available. The list of SNRs also resolved in the TeV band now includes Vela Jr., SN1006, RCW86, IC443, W28, CTB 37B, and others \citep[e.g. database introduced in][]{2012AdSpR..49.1313F}. 
Indeed, the origin of Galactic cosmic rays is one of major motivations for the development of modern experiments in astroparticle physics
such as imaging Cherenkov telescopes measuring very-high-energy gamma rays, like H.E.S.S., MAGIC, VERITAS, and in the future CTA \citep{2013APh....43....3A,2017arXiv170907997C}. 

The total cosmic-ray contribution of an SNR is the time integral of the particle escape rate over the entire evolution of the remnant, which at this time is not well known \citep{2006AdSpR..37.1898P}. In any case, the average particle spectrum is different from the one found in the remnant at any point in time. An accurate particle acceleration modeling at \emph{all} evolutionary phases of SNRs is required for understanding the source spectrum of Galactic cosmic rays. The evolution of an SNR could be subdivided into four major stages: ejecta-dominated, adiabatic phase, post-adiabatic and radiative stage. 
The initial free expansion of ejecta is characterized by the fastest shocks, and as a consequence the most efficient cosmic-ray acceleration; but this stage is rather brief. In the following adiabatic phase, the shock is still fast, and a fair fraction of high-energy cosmic rays produced by the remnant should be accelerated during this stage. In the subsequent post-adiabatic stage,  radiative losses become progressively dynamically important, guiding an SNR to the fully radiative phase. The post-adiabatic shock becomes slow, allowing for efficient recombination in the downstream region. The radiative losses increase the compressibility of matter, which should affect and harden the spectrum of cosmic rays accelerated at this time and will collapse the shocked fluid into a thin dense shell, in which the hadronic gamma-ray production is expected to be very efficient \citep{2015ApJ...806...71L}. 

It is important to stress that the properties of the shock between the second and the fourth stages cannot be adequately described by standard models, neither by adiabatic nor radiative dynamics, implying the necessity to consider an additional intermediate post-adiabatic stage \citep{2005JPhSt...9..364P}. Interestingly, there are evidences that during this stage SNRs become invisible in radio band \citep{2010A&A...509A..34B}. What happens in other wavebands? In particular, do SNRs faint similarly also in leptonic \g-rays? One of the reasons could be a considerable increase of the magnetic field strength after passage through the shock transition region: radiative losses lead to prominent compression of the plasma and thus of the frozen-in (tangential) magnetic field \citep{Petruk-Kuzyo-2016}.      

The post-adiabatic phase is particularly important for SNR-cloud interactions and for the study of high-energy emission coming from accelerated protons. In fact, a strong shock entering the medium with increasing density quickly decelerates, leading to significant radiative losses. A considerable flux of hadronic gamma rays could be expected from such systems as a result of frequent collisions between accelerated protons and target nuclei in the high-density environment either of the post-shock gas (SNR interior) or the pre-shock medium (the cloud). A significant compression of the magnetic field may follow the gas compression, resulting in a suppression of leptonic gamma-ray production on the account of strong synchrotron losses. Cosmic-ray protons will then be the dominant emitters of gamma rays, and (post-adiabatic) SNRs interacting with or situated near the dense clouds should be ideal targets for hadronic gamma-rays observations \citep{2012SSRv..173..369H,2015SSRv..188..187S,2017AIPC.1792b0002G}. 
Indeed, a large fraction of SNRs observed by Fermi gamma-ray observatory \citep{2016ApJS..224....8A} and by Cherenkov telescopes \citep[e.g.][]{2018A&A...612A...3H} are of the middle age and interacting with clouds \citep[e.g.][]{2010ApJ...712.1147J,2017MNRAS.468.2093D}. 

These SNRs are expected to produce also nonthermal X-rays through the synchrotron and/or bremsstrahlung mechanisms, either in case of the shock motion in the cloud \citep{2000ApJ...538..203B} or if the highly-energetic particles diffuse from the SNR shock into the cloud \citep{2009MNRAS.396.1629G}. However, there is still no observational confirmations of such X-rays \citep{2018A&A...612A..32M}.

There are three scenarii for the hadronic \g-rays from SNR-MC complexes. 
A gamma-ray emission from clouds near SNRs can arise if the shock-accelerated protons diffuse towards an adjacent but physically separated cloud \citep{2007ApJ...665L.131G,2009MNRAS.396.1629G,Teletal12b,2015A&A...577A..12F}. One can also envision that the shock moves in a medium which is uniform on large scales but harboring small dense cloudlets and it re-accelerates pre-existing cosmic rays \citep[e.g.][]{2012ApJ...744...71I}. Radiative shock impacting the clumps (`crushing the clouds') enhances the non-thermal emission \citep{1982ApJ...260..625B,2010ApJ...723L.122U,2014ApJ...784L..35T,2016A&A...595A..58C}.In the third scenario, the SNR shock interacts with a large dense cloud. Hydrodynamic (HD) models demonstrate the increase of the gamma-ray flux in such a system \citep{2015ApJ...806...71L}. 

Actually, this third scenario is what we are interested in the present paper. What is missing though are \textit{magneto}-hydrodynamic studies on the shock-cloud interaction. Magnetic field should be quite important party in the highly compressible plasma and is expected to considerably affect the gamma-ray production in systems where an SNR directly interacts with molecular cloud. An additional inevitable factor is that the cloud outskirts and clouds themselves are highly \textit{non-uniform}.\footnote{The `non-uniformity' means the continuous large-scale spatial variation (i.e. comparable to SNR sizes), in contrast to the `inhomogeneity' which marks the multiple discrete small-scale clumps on a background which is uniform on a large scale.}

What is important in any SNR model interacting with a dense cloud is radiation losses of plasma. Effect of energy losses by thermal radiation on the SNR hydrodynamics were studied before \citep{1988ApJ...334..252C,1998ApJ...500..342B}, but very little account was made of the magnetic field role in the SNR evolution. \citet{Petruk-Kuzyo-2016} have demonstrated that the magnetic field is a crucial factor determining dynamics and structure of the post-adiabatic shocks. In particular, a large compression of the transverse magnetic field component leads to a significant transfer of the shock energy into the post-shock magnetic field. The high magnetic pressure in turn prevents the large post-shock density, considerably contrasting the pure hydrodynamic results. 

The medium around SNRs is non-uniform in most cases, especially around those located near the clouds. A number of approximate semi-analytical methods of the shock front dynamics in the medium with density non-uniformities were developed in previous decades \cite[][and references therein]{1995RvMP...67..661B}. A possible way to reconstruct the flow structure downstream of shock in the inter-stellar medium (ISM) with density gradients appears possible in the sector approximation \citep{1999A&A...344..295H}, that allows the authors to analyze the thermal emission of non-spherical SNRs, including their maps. The rise of the numerical power helps to account for the magnetic field non-uniformities as well. In fact, we have explored the internal structure and surface brightness maps of SNR in adiabatic regime, including a full account of the magnetic field, the non-uniform distribution of ambient gas and magnetic field, and oblique orientation of the shock \citep{2007A&A...470..927O,2011A&A...526A.129O}. 

The analysis of the role of nonuniform environment \citep{2001A&A...371..267P}, has allowed to develop a model for the mixed-morphology SNRs; these are SNRs with a shell-like morphology in the radio band but are centrally-filled in thermal X-rays \citep{1998ApJ...503L.167R}. It was argued that these SNRs are physically interacting with molecular clouds, which renders mixed-morphology SNRs likely emitters of hadronic gamma rays \citep{2001A&A...371..267P}. And in fact, 19 SNRs with clear signs of interaction with a molecular cloud are detected in gamma rays, and 10 of them are mixed-morphology SNRs \citep{2010ApJ...712.1147J}.

Not yet studied at all the consequences of non-uniformity at the outskirts of molecular clouds coupled with radiative losses. The recent investigation by \citet{2015ApJ...806...71L} is excellent in its modelling of emission, but does not account the magnetic field and is based on assumptions of a uniform environment. The important questions remain still open: How do the oblique shock orientation, the increasing density and magnetic-field amplitude affect the dynamics of the {post-adiabatic} shock, and hence the flow and magnetic-field profiles, and eventually the spectra of cosmic rays and their emission? 

Since Paper I \citep{Petruk-Kuzyo-2016} shows the negligible role of the parallel MF on the post-adiabatic and radiative SNR evolution, the question is can we explain the evolutionary properties of the oblique post-adiabatic shocks mostly by the perpendicular MF component? This point is answered in Sect.~\ref{rad2:sect3}. One can reasonably expect that a positive density gradient will lead to the shock deceleration, affecting in turn the radiative losses rate. Sect.~\ref{rad2:sect4} deals with the role of ISM density non-uniformities in post-adiabatic evolution, while the medium with the magnetic field gradient is considered in Sect.~\ref{rad2:sect5}. Finally, the model where both non-uniformities are present is studied in Sect.~\ref{rad2:sect6}.

\section{Model and Code}
\label{rad2:sect2}

The system of the time-dependent equations for the ideal MHD is represented by
\begin{equation}
  \frac{\partial}{\partial t} \left(\!\!
  \begin{array}{c}
    \rho \\ \rho \mathbf{v} \\ E \\ \mathbf{B}
  \end{array}
  \!\!\right) + \nabla \cdot \left(\!\!\!
  \begin{array}{c}
    \rho \mathbf{v} \\
    \rho \mathbf{v} \mathbf{v} + \mathbf{B}\mathbf{B} + \mathbf{I}p\rs{tot} \\
    (E+p\rs{tot})\mathbf{v} - \mathbf{B}(\mathbf{v}\cdot\mathbf{B}) \\
    \mathbf{v}\mathbf{B} - \mathbf{B}\mathbf{v} \\
  \end{array} 
  \!\!\!\right)^\mathrm{T} \!\! = \left(\!\!
  \begin{array}{c}
    0 \\ 0 \\ L \\ 0
  \end{array}
  \!\!\right)
  \label{rad2:eq1}
\end{equation}
where $\rho$ is the density, $\rho \mathbf{v}$ the momentum density, $\mathbf{v}$ the flow velocity, $p\rs{tot}$ the total (thermal $p$ and magnetic $p\rs{B}$) pressure, $\mathbf{I}$ the unit vector, $\mathbf{B}$ the MF strength, $L$ represents the radiative losses of plasma, and $E$ is the total energy density. Also the ideal gas equation of state is considered with $\gamma=5/3$. The total energy density is a sum of the thermal, kinetic and magnetic components:
\begin{equation}
  E=\frac{p}{\gamma-1}+\frac{\rho v^2}{2}+\frac{B^2}{2}.
\label{rad2:energy_eq}
\end{equation}
Additionally, the divergence-free condition $\nabla \cdot \mathbf{B}=0$ holds. 

\begin{figure}
  \centering 
  \includegraphics[width=8.7truecm]{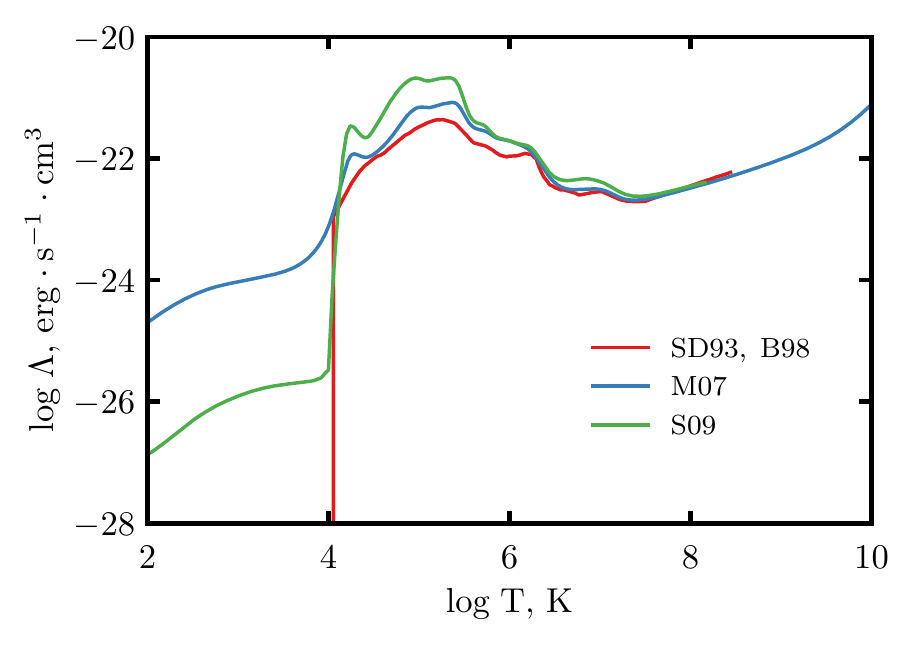}
  \caption{Temperature dependence of the cooling coefficient.  
  Abbreviations refer: SD93 to  \citet{1993ApJS...88..253S}; B98 to \citet{1998ApJ...500..342B}; 
  M07 to \citet[][{\sf PLUTO} v.4.2]{Mignone2007}; 
  S09 to \citet{2009A&A...508..751S}. 
  }
  \label{rad2:fig_cooling}
\end{figure}

In the present paper, we are interested in the late times and regions near the SNR shell. The ejecta is very deeply inside SNR at these stages and does not affect anything around the shock. Thus, all we require from the explosion is to generate the strong shock. Therefore, the initial conditions correspond to the point explosion (without ejecta). The entire explosion energy $E\rs{o}$ is placed as the thermal component, uniformly in the small volume (15 grid zones) at the grid origin. We solve the system (\ref{rad2:eq1}) numerically in one dimension (radial in spherical coordinates). 1-D simulations allow us to have a very high spatial resolution and as a consequence to reveal thin features developing behind radiative shocks. In fact, we adopted a uniform grid of $60\, 000$ zones which spans over 30 pc.
The 1-D simulations in the Paper I and in the present study are the reference simulations in order to interpret the forthcoming 3-D MHD simulations. 3-D MHD numerical task cannot be performed with so high resolution and may not therefore resolve some details behind the shocks modified by the radiative losses.

In the case of evolved SNRs, the thermal conduction may play a role also in contrasting the radiative cooling (e.g. \citealt{2008MNRAS.386..642B}). However, in the presence of a tangential component of the ambient magnetic field, as in the case of oblique shocks, the heat conduction can be largely reduced, depending on the magnetic field strength (e.g.  \citealt{2008ApJ...678..274O}). In our simulations, we assume that the magnetic field is intense enough to limit the effects of thermal conduction which is, therefore, neglected in our calculations.

As in the Paper I, we make use of {\sf PLUTO} MHD code \citep{Mignone2007,Mignone2012}
and consider the same physical parameters for the problem setup: supernova explosion energy $E\rs{o} = 10^{51} \un{erg}$, ISM hydrogen number density around the pre-supernova location $n\rs{Ho} = 0.84 \un{cm^{-3}}$ and temperature $T\rs{o} = 10^4\un{K}$.
We keep the temperature of the ambient medium constant. The sound speed, which is proportional to $T\rs{o}^{1/2}$, is also constant in ISM; its value is $15 \un{km/s}$. In our simulations, the shock speed appears to be larger than $50\un{km/s}$  and the Mach number exceeds $3$ even at the latest times. 
We refer readers to Paper I for description of the computational schemes and other numerical setup details.

Plasma energy losses due to the thermal radiation are given by the term
\begin{equation}
 L = -n\rs{e} n\rs{H} \Lambda(T),
 \label{rad2:radlossdef}
\end{equation} 
where $n\rs{e}$ and $n\rs{H}$ are the electron and hydrogen number densities, $T$ is the plasma temperature and $\Lambda(T)$ is the cooling coefficient.
In Paper I, we did not use the default cooling data provided by the developers of {\sf PLUTO} (blue line on Fig.~\ref{rad2:fig_cooling}). Instead, we have adopted the cooling curve from \citet{1993ApJS...88..253S} (red line on Fig.~\ref{rad2:fig_cooling}) in order to have a possibility to directly compare our results to findings of \citet{1998ApJ...500..342B}. In the present paper, we keep using the same cooling coefficient as in Paper I but for $T> 10^{4.1}\un{K}$ only. As to the lower temperatures, in answer to a recent note by \citet{2016MNRAS.459.2188B}, we extend the cooling curve by the lower-temperature data given by \citet{2009A&A...508..751S} (i.e. we switch to the green line for $T\leq 10^{4.1}\un{K}$, see   Fig.~\ref{rad2:fig_cooling}). 
The contribution to the low-temperature part of cooling curve is primarily a result of excitation of singly charged ions with the thermal electrons and collisions of hydrogen atoms with electrons \citep{2009A&A...508..751S}. Effects caused by this part of $\Lambda(T)$ will be discussed in Sect.~\ref{rad2:sect3}.

\section{Oblique Shocks}
\label{rad2:sect3}

The properties of the strong shocks propagating in the magnetized medium are characterized by the obliquity angle $\Theta\rs{o}$. It is defined as the angle between $\mathbf{B}\rs{o}$ and the normal to the shock surface (the index `o' marks the pre-shock values). The role of the magnetic field on the post-adiabatic evolution of the parallel ($\Theta\rs{o}=0$) and perpendicular ($\Theta\rs{o}=90^\circ$) shocks in the uniform medium was studied in the Paper I. In current section we consider shocks of the intermediate obliquities moving in the medium with uniform density and MF distributions. 

We expected and have actually found that properties of the shocks with the intermediate obliquities are `in-between' scenarios of parallel and perpendicular cases.

One of the simulation accuracy metrics is the sum of all the energy components (plasma thermal energy $E\rs{th}$, kinetic energy of the flow $E\rs{kin}$, the energy in magnetic field downstream $E\rs{B}$ and the energy radiated away due to the plasma losses $E\rs{rad}$). The total energy in the whole system is preserved with the very high degree of accuracy during the whole integration time, regardless of the shock obliquity.
The sums of energy components $E\rs{th}+E\rs{rad}$ and $E\rs{kin}+E\rs{B}$ remain almost constant during the evolution, as it is demonstrated on Fig.~5 in Paper I for the perpendicular shock. Interestingly, the plot looks almost the same for any $\Theta\rs{o}$: the fractions  $(E\rs{th}+E\rs{rad})/E\rs{o}\approx 0.7$ and $(E\rs{kin}+E\rs{B})/E\rs{o}\approx 0.3$. 

\begin{figure}
  \centering 
  \includegraphics[width=8.7truecm]{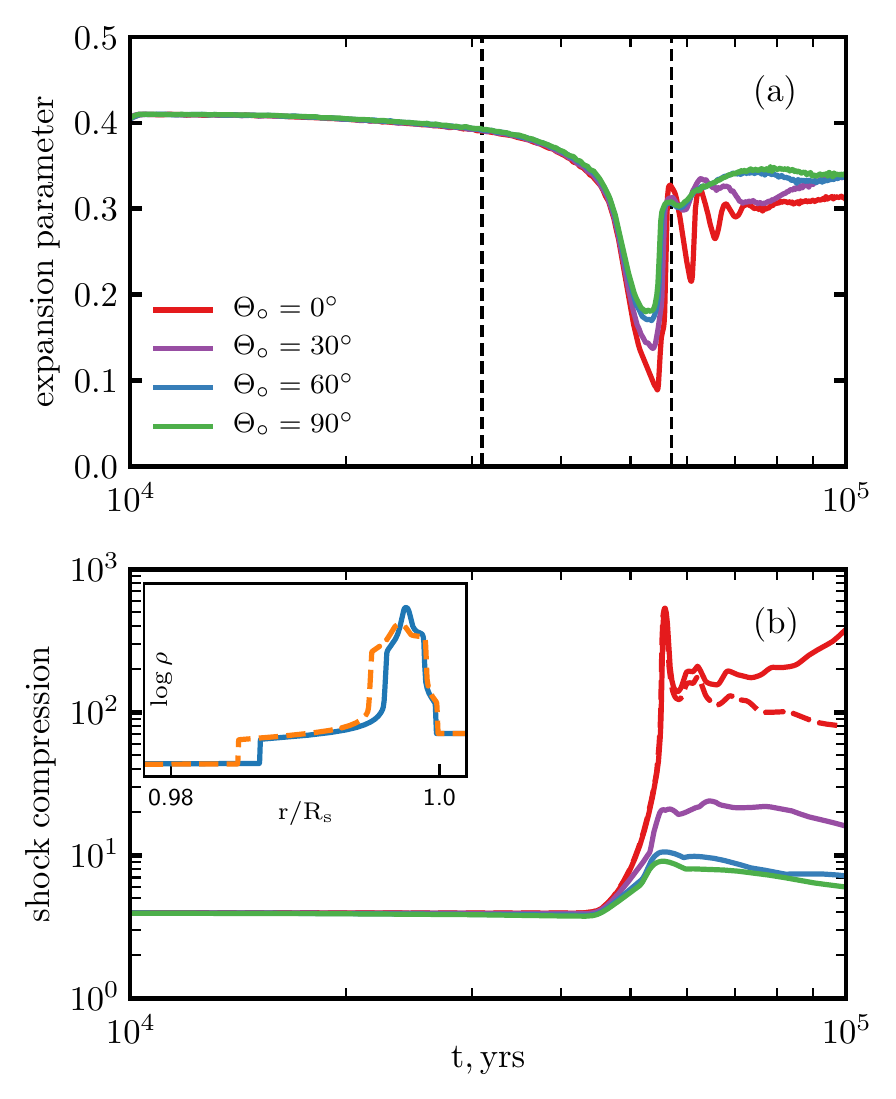}
  \caption{Temporal evolution of the expansion parameter $m$ ({\bf a}) 
  and the shock compression factor $\rho\rs{max}/\rho\rs{o}$ ({\bf b}), 
  for several obliquity angles $\Theta\rs{o}$ and MF strength $B\rs{o} = 10 \un{\mu G}$. 
  The vertical dashed lines on the plot {\bf a} indicate $t\rs{tr}=3.1\E{4}\un{yrs}$ and $t\rs{sf}=5.7\E{4}\un{yrs}$. 
  The dashed lines on the plot {\bf b} correspond to the model with the cooling curve $\Lambda(T)$ from Paper I  which differs from the cooling curve in the present paper by the low-temperature part (below $10^{4.1}\un{K}$, see Sect.~\ref{rad2:sect2}). 
  The inset plot shows the post-shock density profiles for $t = 8\E{4}\un{yrs}$ and demonstrates the differences due to these cooling functions.
   }
  \label{rad2:fig_m_rho_max_oblique}
\end{figure}

The deceleration (or expansion) parameter $m=-d \ln R/d \ln t$ is an important marker of the shock evolutionary stage (Fig.~\ref{rad2:fig_m_rho_max_oblique}a). In particular, $m=0.4$ corresponds to the Sedov phase. It starts to deviate from this value around the `transition' time  $t\rs{tr}$ which marks the end of the adiabatic and the beginning of the post-adiabatic stage. The next fully radiative phase starts around the `shell-formation' time $t\rs{sf}$ \citep[for details see][and references therein]{2005JPhSt...9..364P}. The time dependence of $m$ for the fully radiative shocks, in the pure HD model, is given by the equation (23) in \citet{2004A&A...419..419B}: the value of $m$ reaches a local maximum at some time after $t\rs{sf}$ and then, at late radiative stage, tends to the asymptotic value $2/7$ found by \citet{1977ApJ...218..148M}. 

Fig.~\ref{rad2:fig_m_rho_max_oblique}a shows the behavior of the deceleration parameter $m$ for MHD shocks of different obliquities (cf. Fig.4b in Paper I which represents the same dependence for different values of $B\rs{o}$). The red line (parallel shock) coincides with the $B\rs{o}=0$ scenario (since the parallel MF is not efficient in modifying the shock dynamics). With the increase of the obliquity, the influence of MF is more prominent. The reason is in the increase of the tangential MF component which, being compressed by the shock, becomes dynamically important.

An important conclusion follows from Fig.~\ref{rad2:fig_m_rho_max_oblique}a. Namely, the shock obliquity does not alter the temporal development of the shock from the adiabatic to the radiative phase. More precisely: in the same system, the minimum and the first maximum in the evolution of $m$ (which marks prominent physical changes in the system) occur at the same time for shocks of any obliquity.

\begin{figure*}
  \centering 
  \includegraphics[width=13.5truecm]{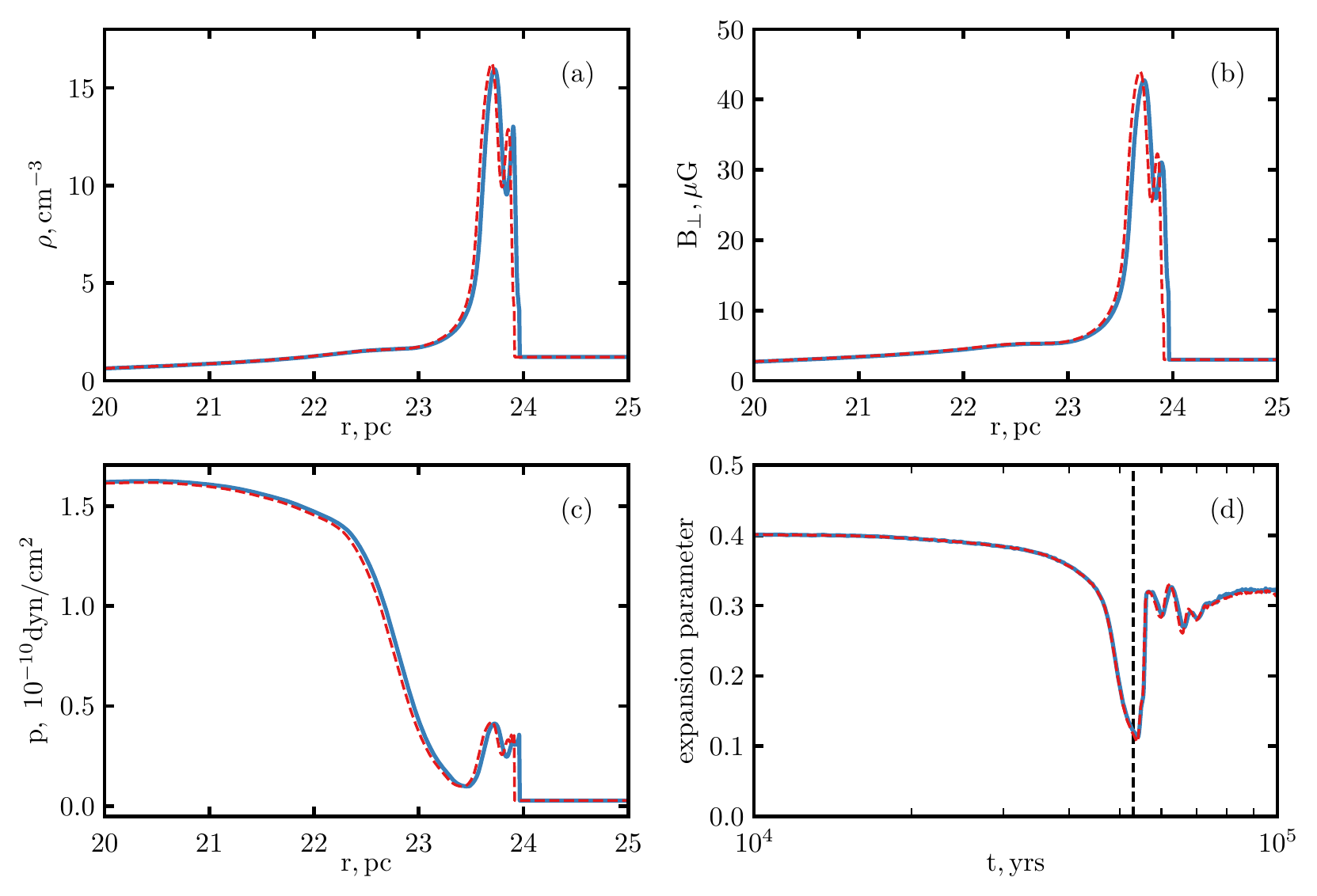}
  \caption{Post-shock structure of $\rho$, $B\rs{\perp}$ and $P$ ({\bf a-c}) 
     behind the shock moving in ISM with $B\rs{oa} = 10\un{\mu G}$ and the obliquity 
     $\Theta\rs{oa} = \arcsin (3/10) \approx 18^\circ$ (solid blue line) and behind 
     the perpendicular shock with $B\rs{ob} = 3\un{\mu G}$ (dashed red line). 
     The profiles were taken at the time $t=53\,000\un{yrs}$,
     which is marked by the vertical dashed line on the plot ({\bf d}) where
     the expansion parameter $m$ (characterizing evolutionary stage) is shown for both models.
  }
  \label{rad2:fig_asin_3_10}        
\end{figure*}

The role of MF obliquity is clearly visible on the plot for the density jump on the shock (Fig.~\ref{rad2:fig_m_rho_max_oblique}b). Note that the shock compression shown here is calculated as $\rho\rs{max}/\rho\rs{o}$ where $\rho\rs{max}=\rho\rs{s}$, i.e. the immediately post-shock value, for the adiabatic and the early post-adiabatic stages while it is the maximum density in the thin radiative shell for the rest of the post-adiabatic phase and for the radiative shocks (that is when this maximum is larger than the immediate post-shock value; note that it is still very close to the shock, see e.g. blue line on the inset in Fig.~\ref{rad2:fig_m_rho_max_oblique}b or the highest peak on Fig.~\ref{rad2:fig_asin_3_10}a).

It was shown in the Paper I that the flow with radiative losses is affected by the tangential MF. In fact, MF is compressed at the perpendicular shock. The higher the radiative losses the larger extent the shock is compressed to giving rise to the magnetic pressure which effectively prevents the creation of the thin high density radiative shell.\footnote{A similar effect may be due to efficiently accelerated cosmic rays: their pressure can work against the radiative compression \citep{2015ApJ...806...71L}.} In particular, for plain HD simulations the ratio of the post- to the pre-shock densities could reach a few hundreds while the presence of $B\rs{o}>3\un{\mu G}$ limits it to $\rho\rs{s}/\rho\rs{o}<10$ \citep[Fig.~8 in][]{Petruk-Kuzyo-2016}. The parallel shock does not compress MF and thus the ratio $\rho\rs{s}/\rho\rs{o}$ is as high as in the HD case (red line on Fig.~\ref{rad2:fig_m_rho_max_oblique}b). Therefore, as Fig.~\ref{rad2:fig_m_rho_max_oblique}b demonstrates, the more oblique is the shock (i.e. the larger the tangential MF component) the smaller is extent to which the plasma can be compressed downstream the radiative shock. 

One can compare Fig.~\ref{rad2:fig_m_rho_max_oblique}b with Fig.8 in Paper I where the time evolution of the shock compression is shown for the perpendicular shocks with different MF strength values. The plots are similar: in both cases, at the end of the post-adiabatic phase and during the radiative stage, the ratio $\rho\rs{max}/\rho\rs{o}$ is reduced with the larger tangential MF. So, there is some degree of ambiguity between the obliquity and MF strength at the perpendicular shock. 

What kind of influence on the flow we may expect from the magnetic field? The radial MF has no effect and the major impact on the flow with the radiative shock originates from the tangential MF component (Paper I). 
We can always split the oblique MF to the radial and tangential components and largely expect that the overall effect from MF of the flow behind the oblique shock is from the tangential MF. Our simulations confirm this.
One can ask the question whether the MHD parameter profiles behind the oblique shock may be approximated by the profiles behind a `corresponding' perpendicular shock. In fact, this is possible. Fig.~\ref{rad2:fig_asin_3_10} demonstrates that the flow structure downstream of the oblique shock with $\Theta\rs{oa}$ in MF $B\rs{oa}$ is quite accurately approximated by the structure of the pure perpendicular shock in ambient MF with a smaller strength $B\rs{ob}$, which has to be just $B\rs{ob}=B\rs{oa}\sin(\Theta\rs{oa})$.

\begin{figure*}
  \centering 
  \includegraphics[width=13.5truecm]{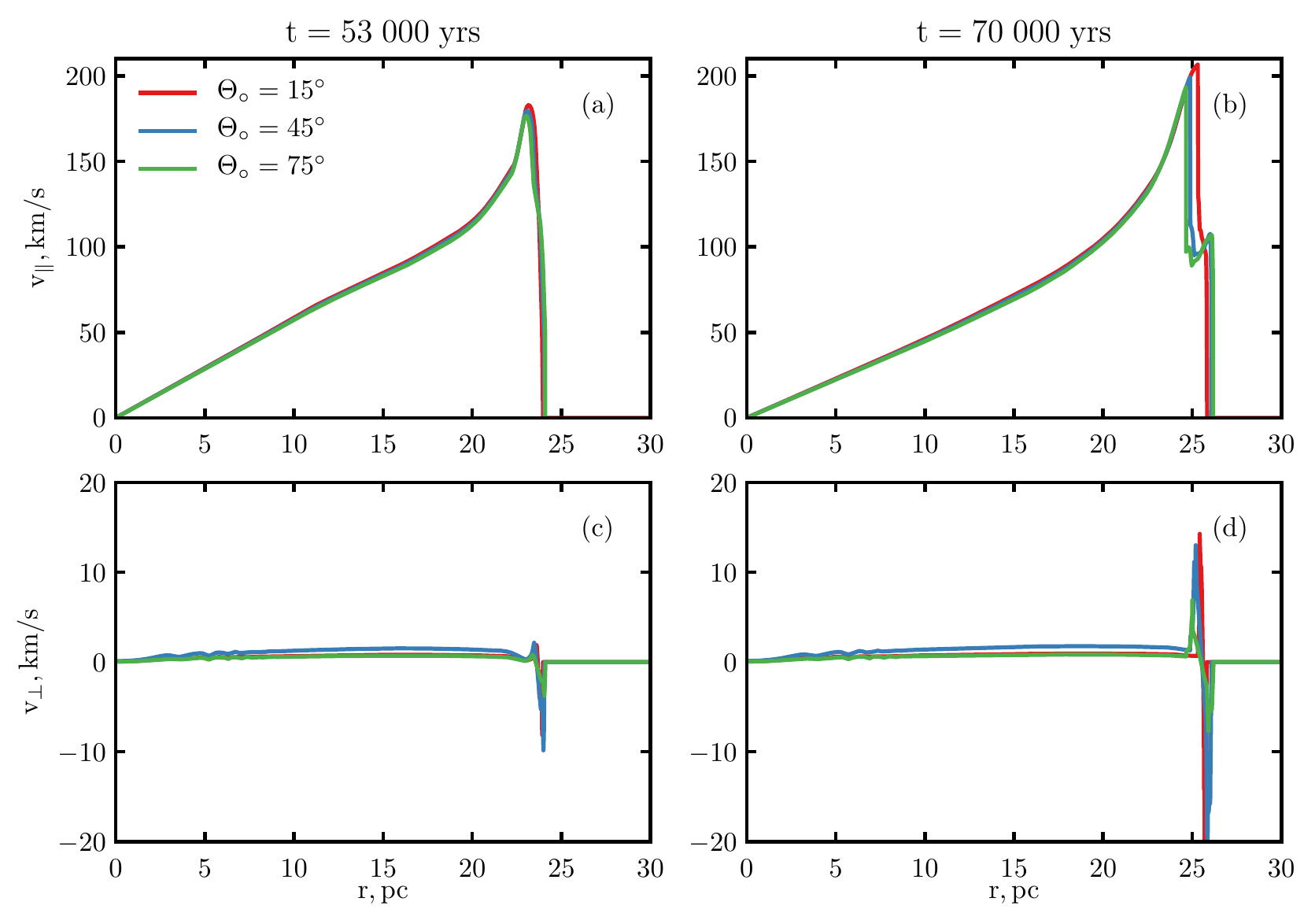}
  \caption{Radial profiles of the radial $v\rs{\|}$ (upper plots) and 
  tangential $v\rs{\perp}$ (bottom plots) velocity components 
    for $t = 53\;000\un{yrs}$ (left) and $t = 70\;000\un{yrs}$ (right). 
    $B\rs{o} = 10\un{\mu G}$. 
  }
\label{rad2:fig_velocity_oblique}
\end{figure*}

Our simulations (Fig.~\ref{rad2:fig_velocity_oblique}) as well as the equation for the momentum density in spherical coordinates 
\begin{equation}
  \dfrac{\partial \rho v\rs{\perp}}{\partial t} + 
  \nabla\cdot(\rho v\rs{\perp}\mathbf{v} - B\rs{\perp}\mathbf{B}) 
  = {-\dfrac{\rho v\rs{\perp}v\rs{\|} - B\rs{\perp}B\rs{\|}}{r}}
  \label{rad2:eq_v_theta}
\end{equation}
show another physical effect for the oblique shocks which is rather counterintuitive in the case of the 1-D simulations. Namely, there exists the non-zero component of velocity $v\rs{\perp}$, perpendicular to the radial direction. This component is only present right behind the oblique shock and vanishes in the boundary cases of the parallel or perpendicular shocks. Its magnitude may reach up to $10\%$ of the radial velocity $v\rs{\|}$ at the radiative stage (Fig.~\ref{rad2:fig_velocity_oblique}). 
The presence of $v\rs{\perp}$ can be interpreted in terms of the post-shock motion of the plasma in the transverse directions. 

\begin{figure*}
  \centering 
  \includegraphics[width=15.6truecm]{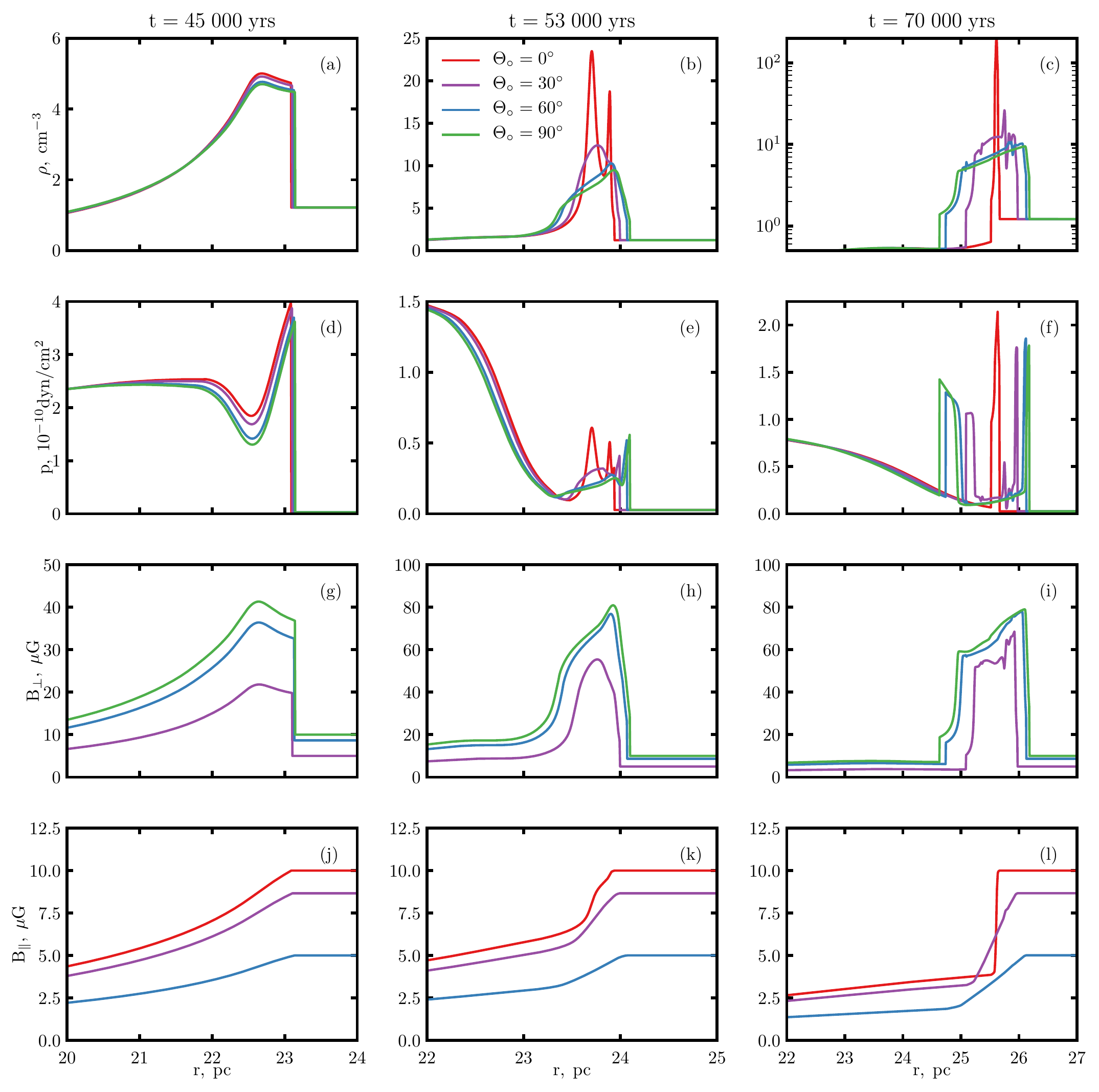}
  \caption{Post-shock radial profiles of the density (1st row), thermal pressure (2nd row), tangential (3rd row) and radial (4th row) MF components. Different stages of SNR evolution (columns) and a few obliquity angles $\Theta\rs{o}$ (colors) are presented. Ambient MF $B\rs{o} = 10\un{\mu G}$.
     (MF values before the shock are not the same for lines of different colors because they represent the MF components $B\rs{o\|}=B\rs{o}\cos\Theta\rs{o}$ and $B\rs{o\perp}=B\rs{o}\sin\Theta\rs{o}$ at different obliquities.)
  }
  \label{rad2:fig_profiles_oblique}        
\end{figure*}
\begin{figure*}
  \centering 
  \includegraphics[width=0.98\textwidth]{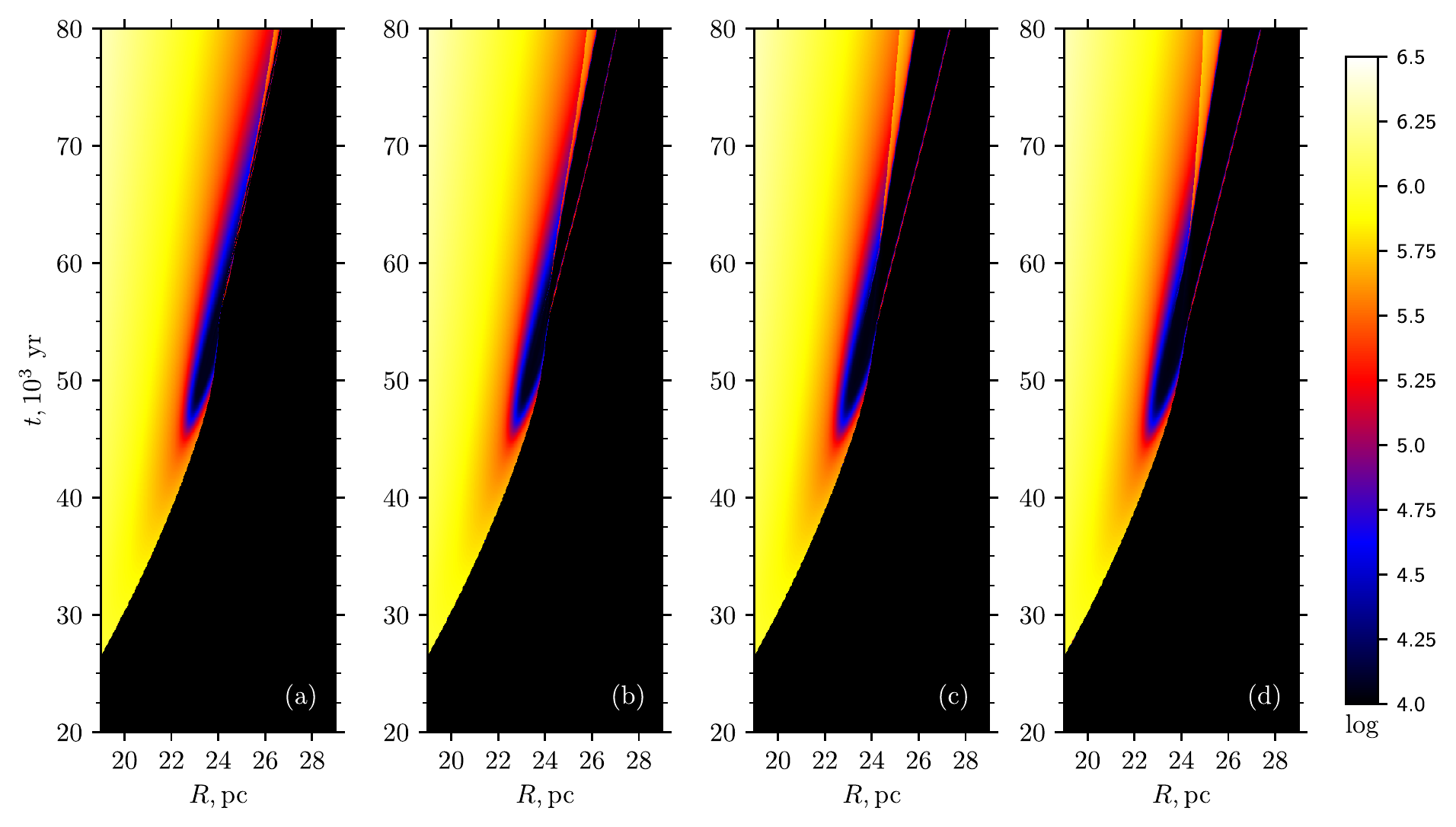}
  \caption{Time dependence of the temperature distribution
    behind the shock for obliquity angles $\Theta\rs{o} = 0^\circ$ ({\bf a}), 
    $\Theta\rs{o} = 30^\circ$ ({\bf b}),  $\Theta\rs{o} = 60^\circ$ ({\bf c})
    and  $\Theta\rs{o} = 90^\circ$ ({\bf d}). Ambient MF strength is $B\rs{o} = 10\un{\mu G}$.
  }
  \label{rad2:fig_oblique_space_time_T}
\end{figure*}
\begin{figure*}
  \centering 
  \includegraphics[width=0.98\textwidth]{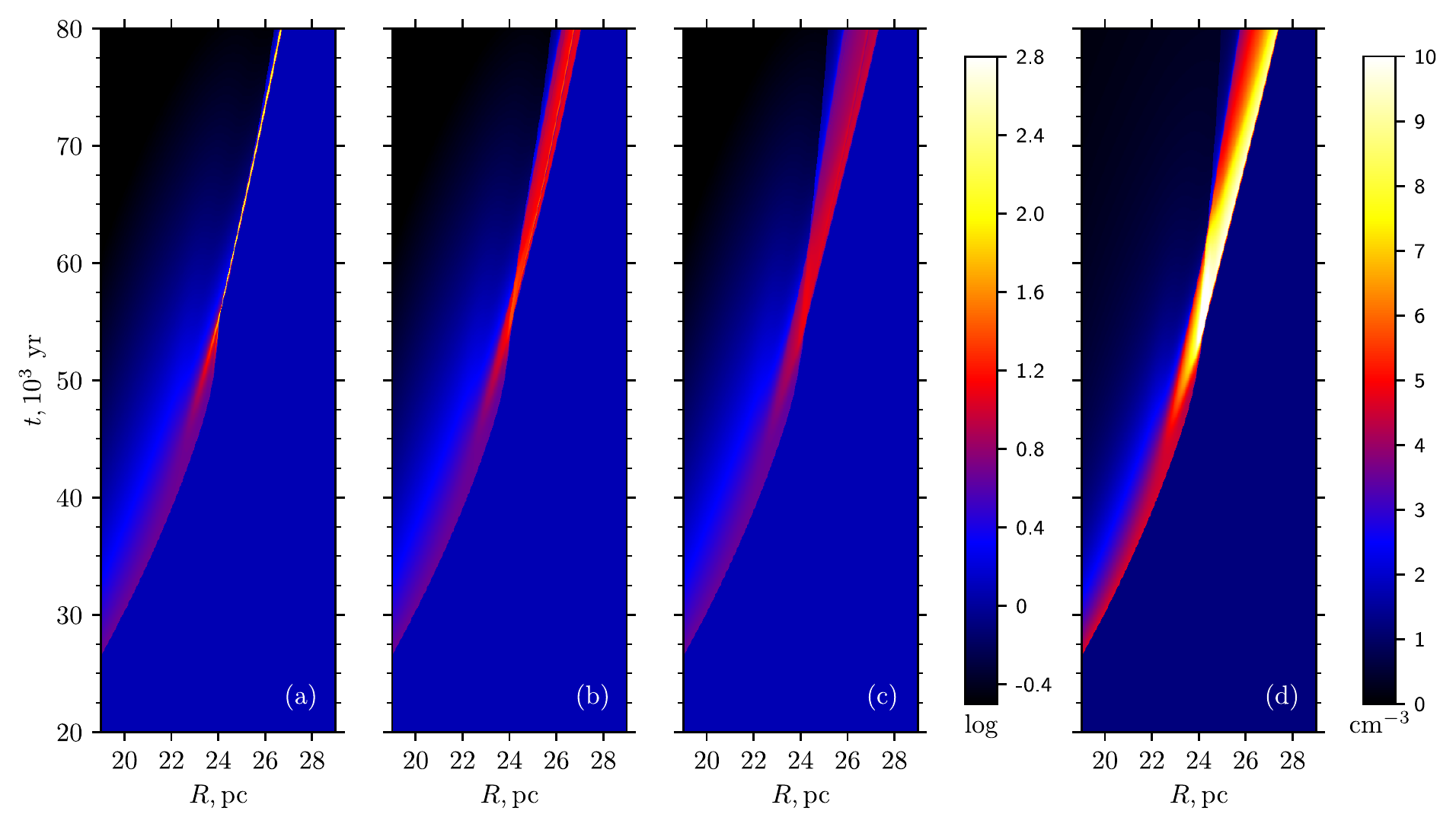}
  \caption{The same as Fig.~\ref{rad2:fig_oblique_space_time_T} for the density distribution. Note that the color bar is logarithmic for plots ({\bf a})-({\bf c}) and linear for ({\bf d}).
  }
  \label{rad2:fig_oblique_space_time_rho}
\end{figure*}
\begin{figure*}
  \centering 
  \includegraphics[width=7.8truecm]{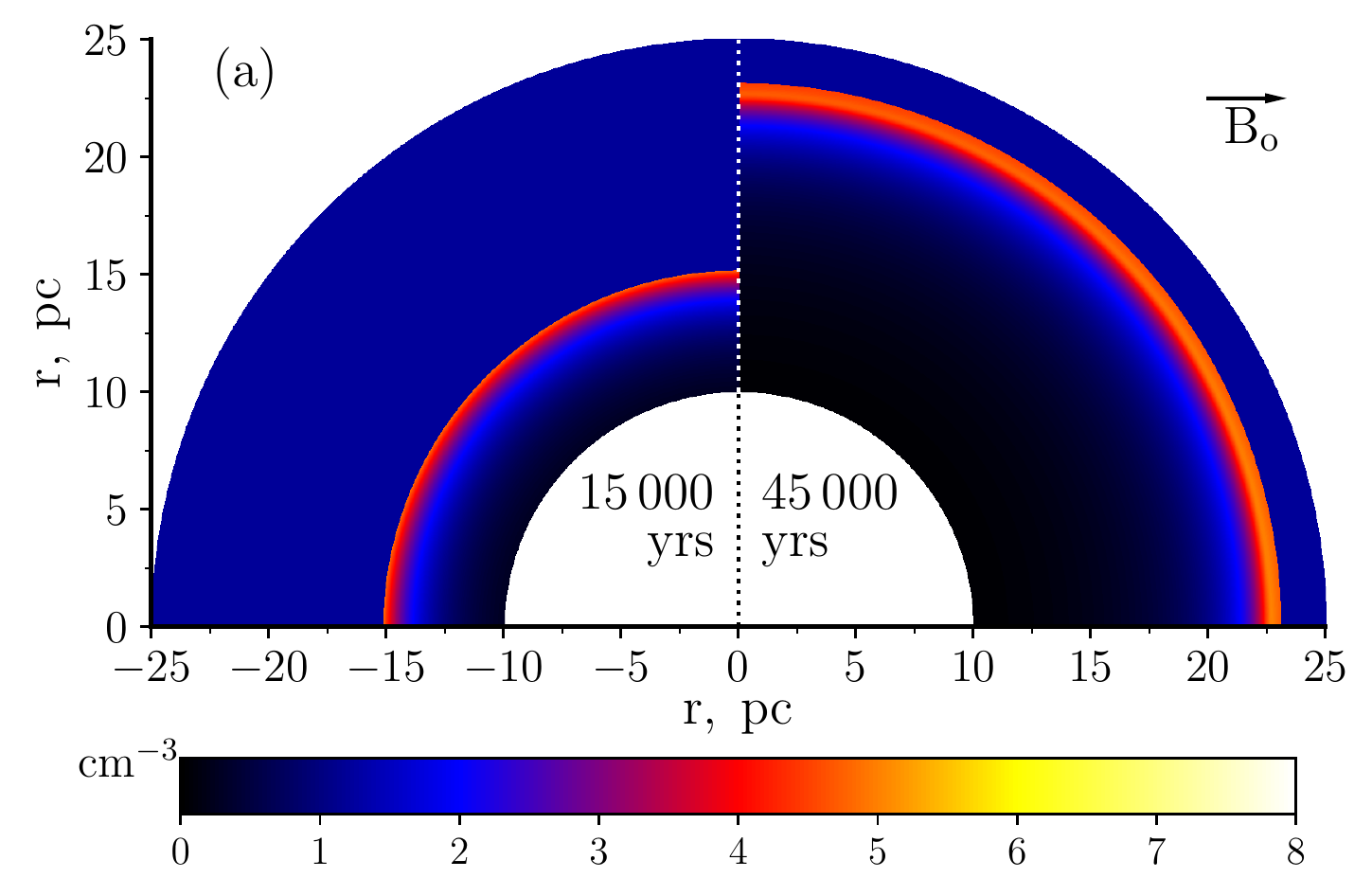}
  \includegraphics[width=7.8truecm]{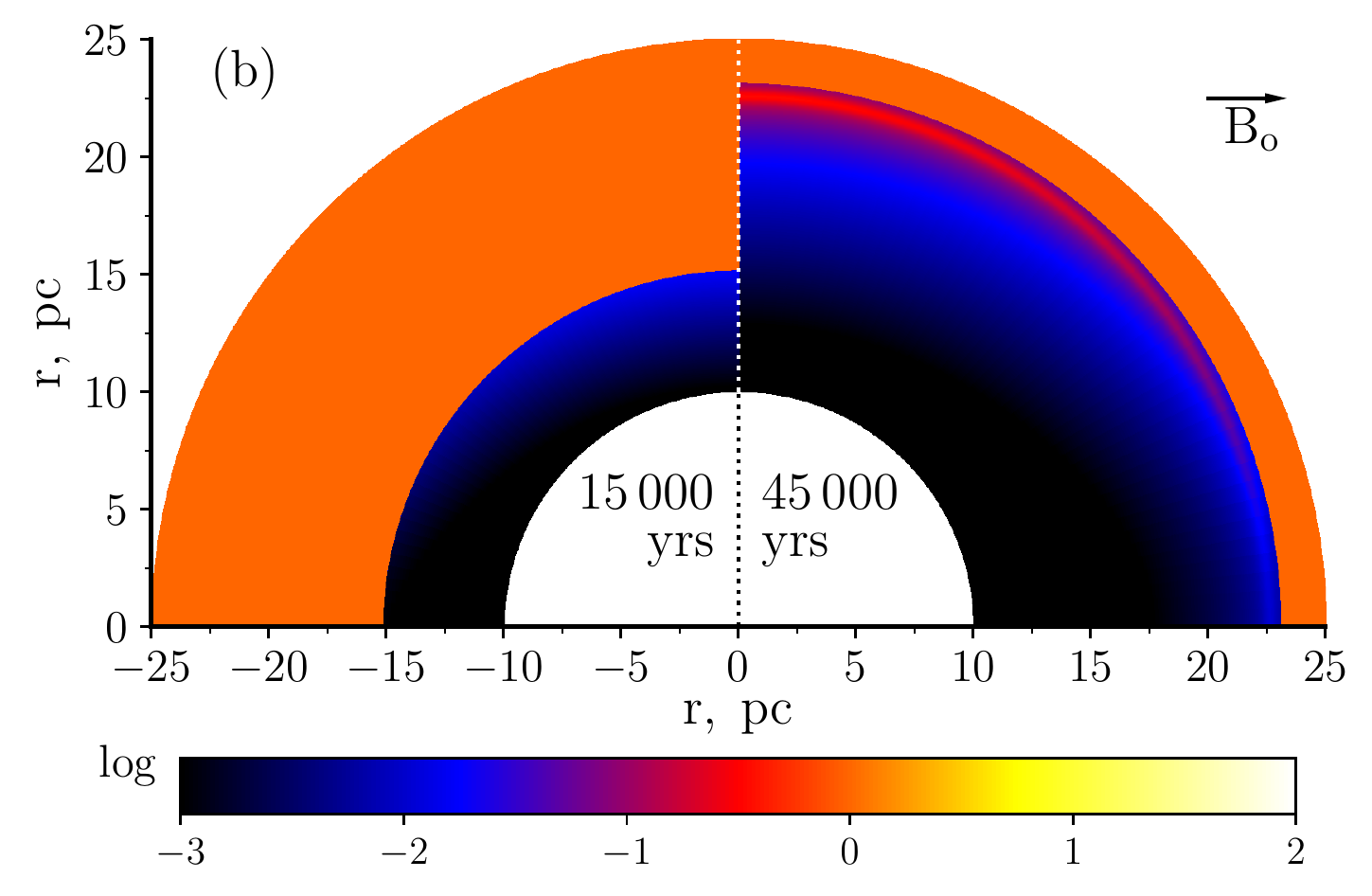}\\
  \includegraphics[width=7.8truecm]{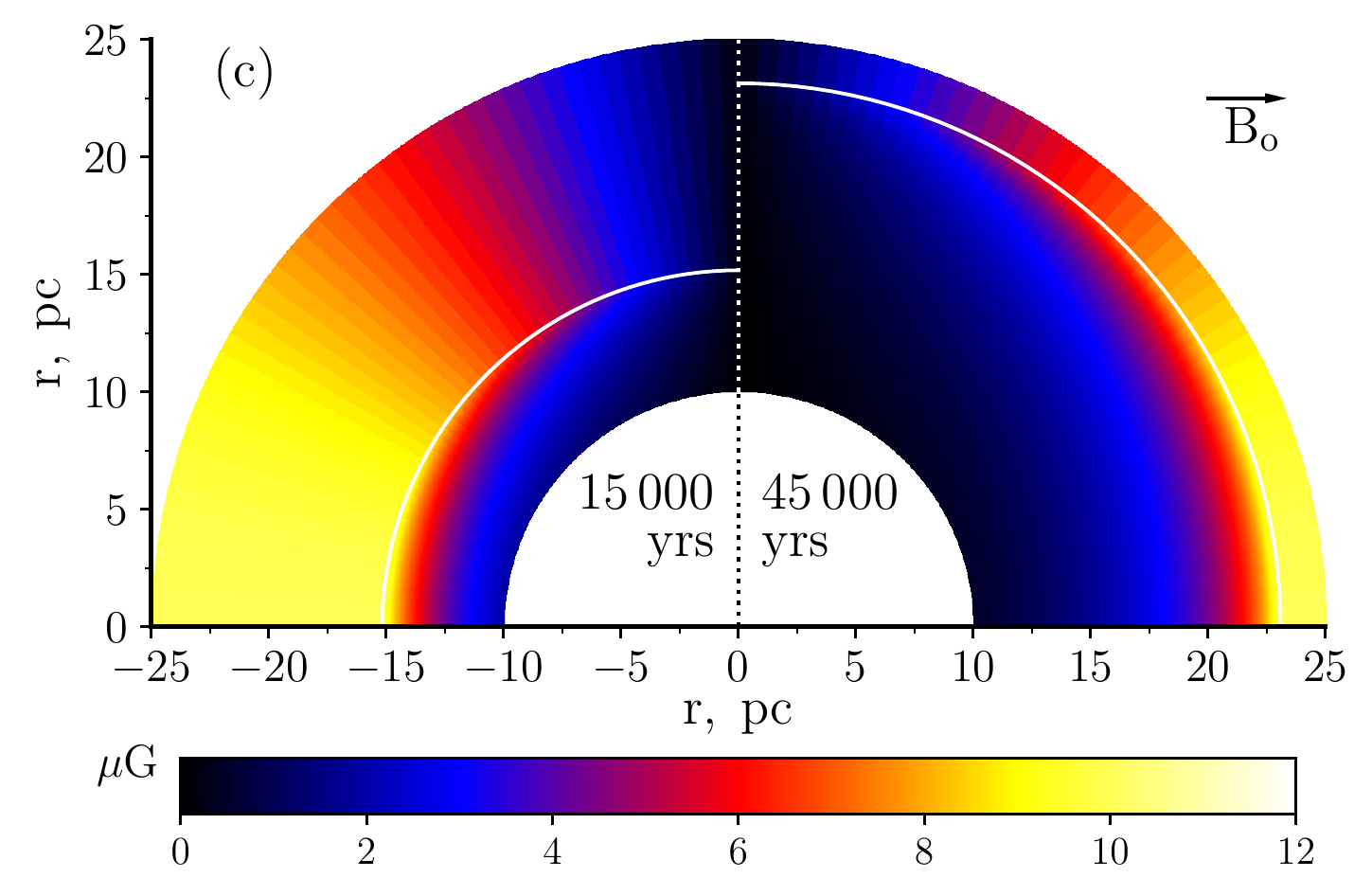}
  \includegraphics[width=7.8truecm]{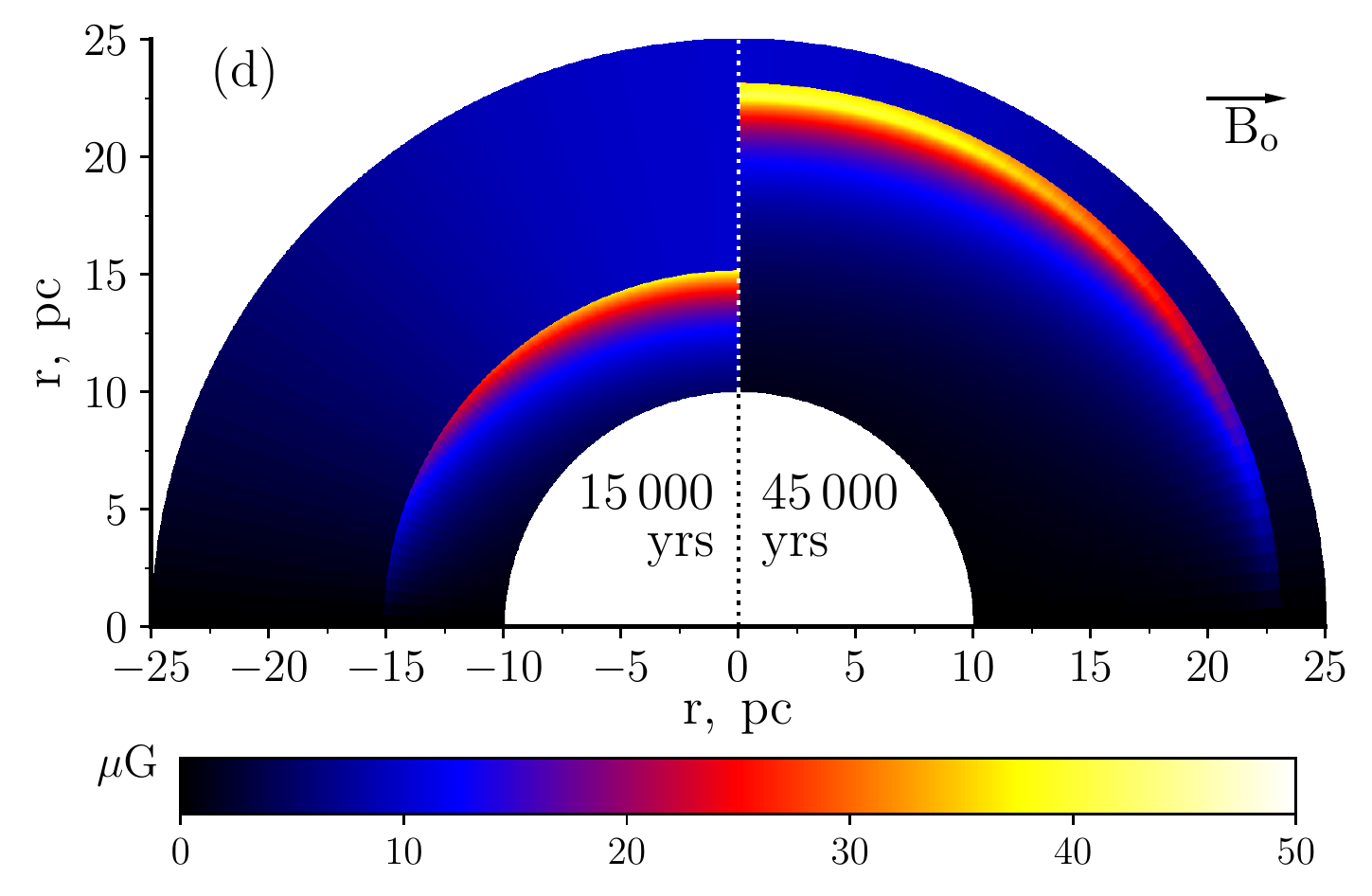}  
  \caption{The angular variations of the number density ({\bf a}), $\beta$ ({\bf b}), 
  radial ({\bf c}) and tangential ({\bf d}) MF components 
  for 15 000 yrs (left half of each panel) and 45 000 yrs (right half of each panel). 
  Ambient MF of the strength  $B\rs{o} = 10\un{\mu G}$ is directed along the horizontal axis.
  (The internal parts were made white on these plots on purpose, in order to emphasize by the color-scale the structures near the shock.)
  }
  \label{rad2:fig_angular1_oblique}
\end{figure*}
\begin{figure*}
  \centering 
  \includegraphics[width=7.8truecm]{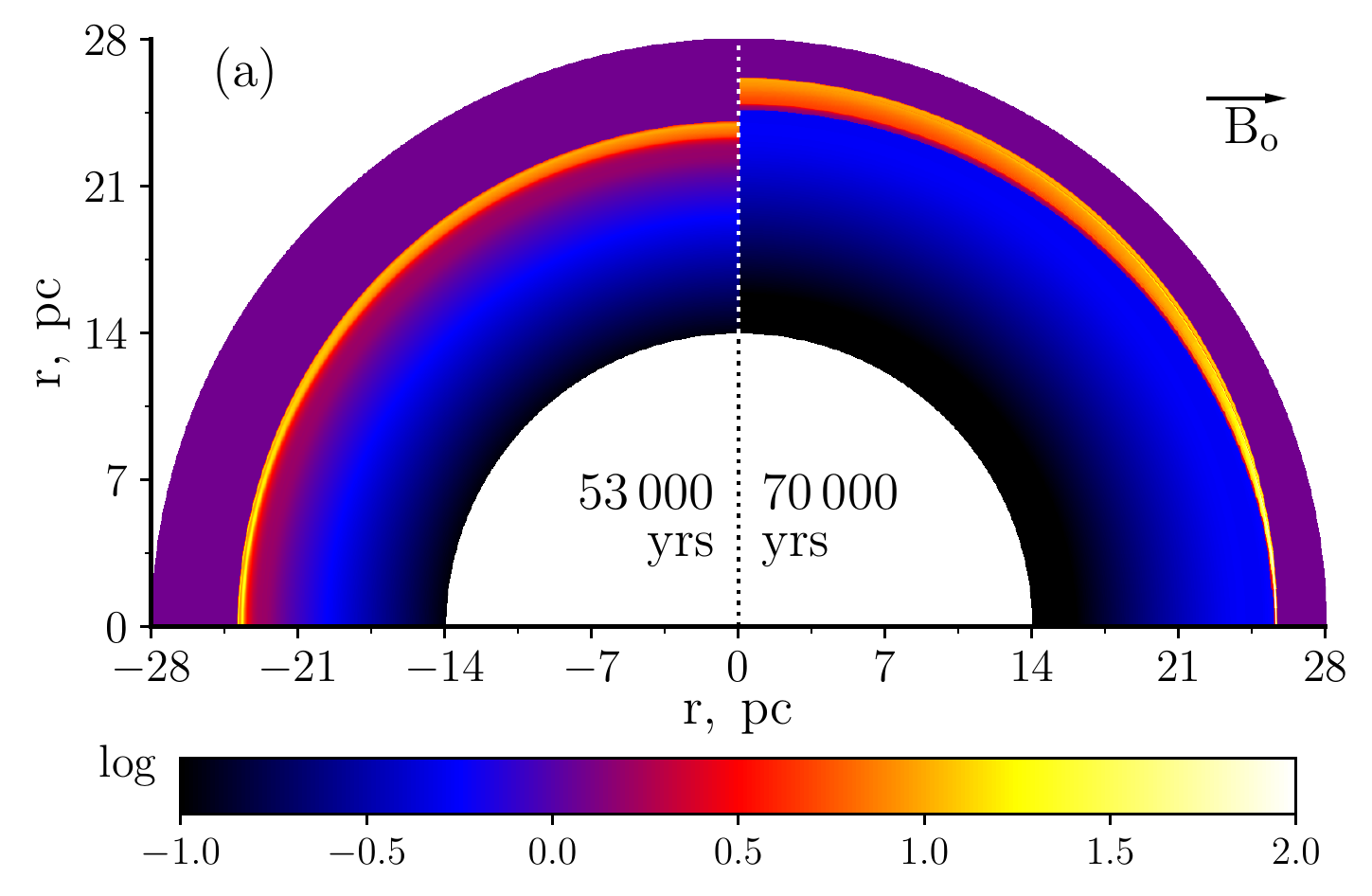}
  \includegraphics[width=7.8truecm]{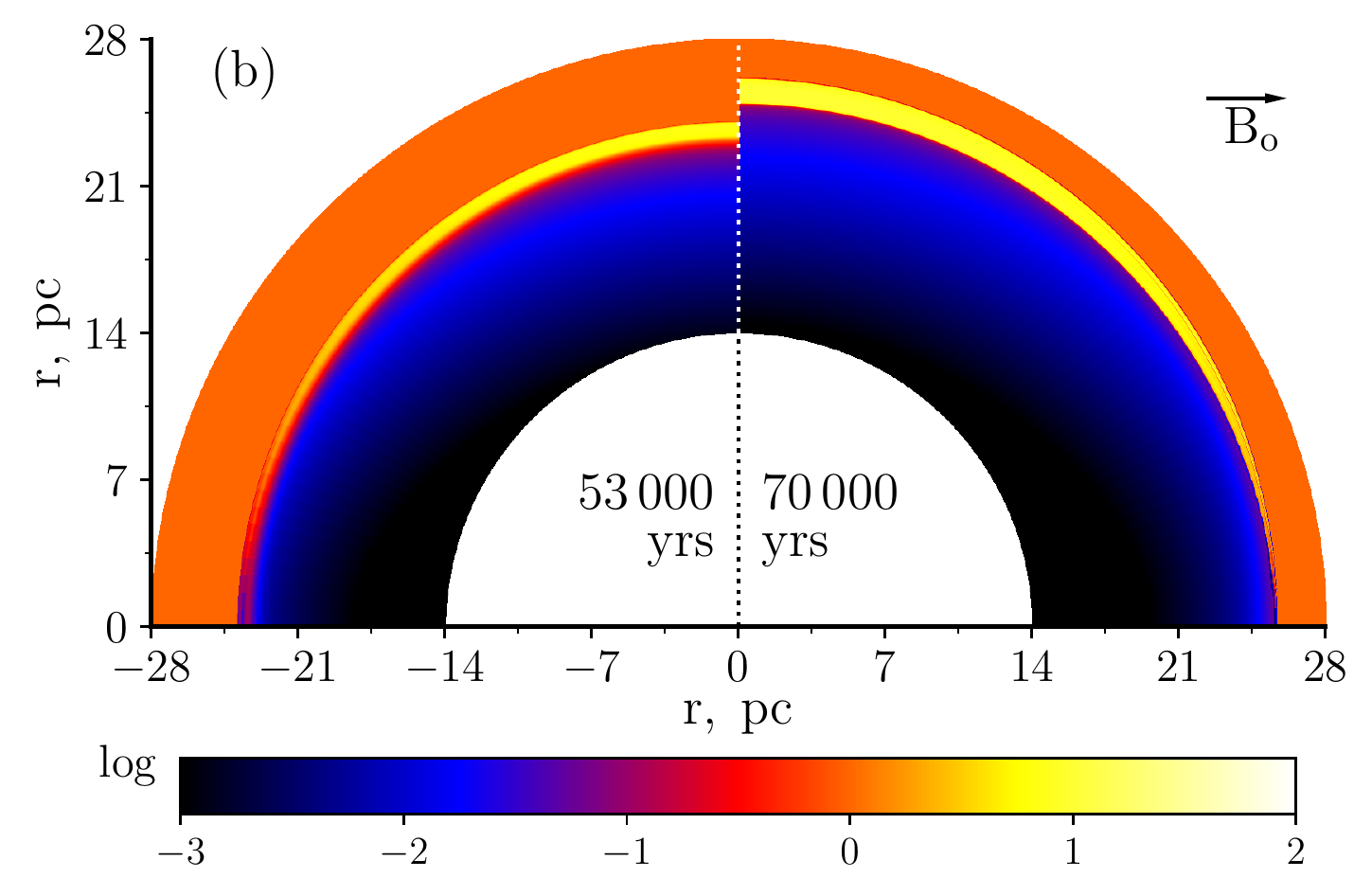}\\
  \includegraphics[width=7.8truecm]{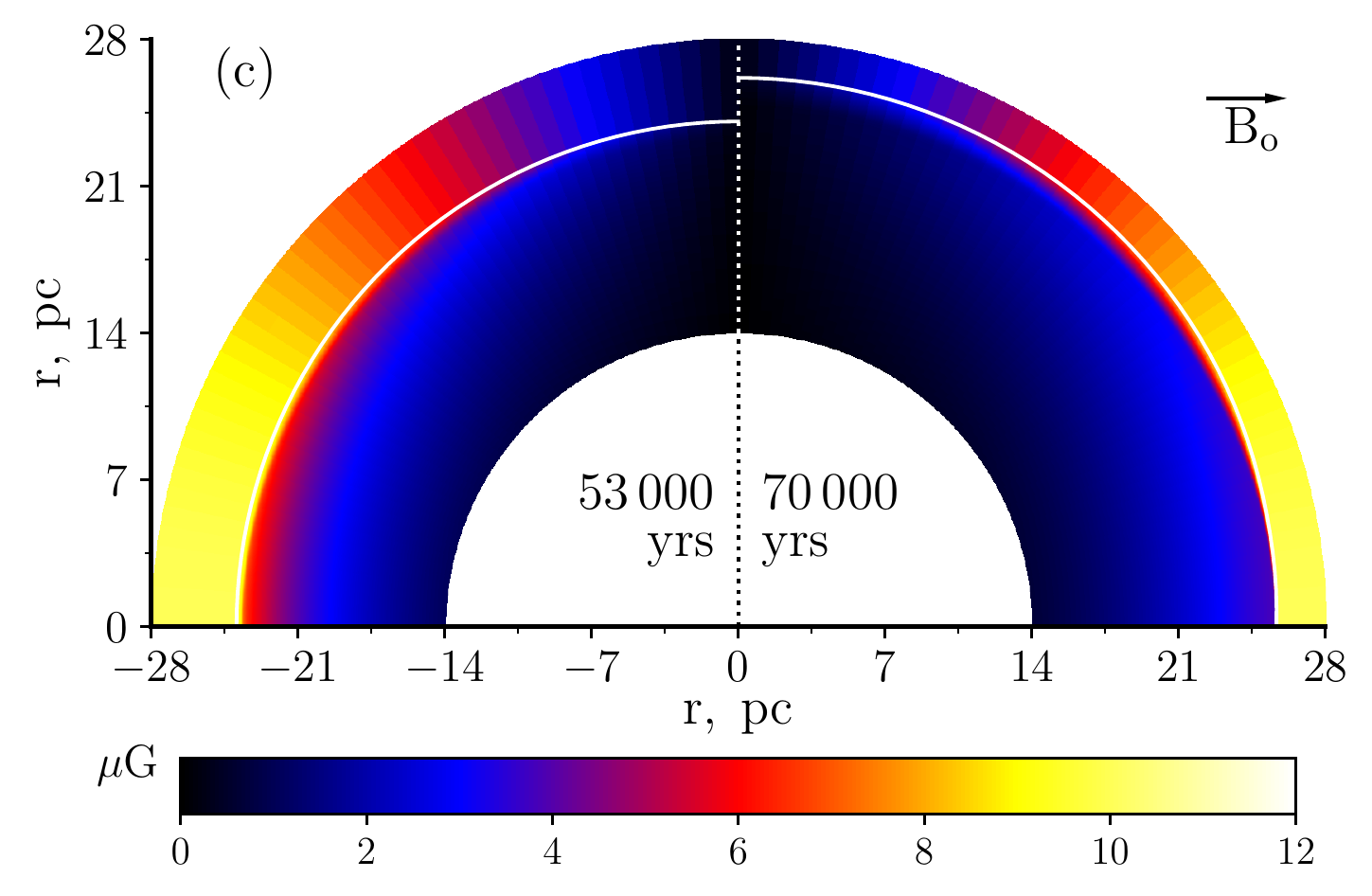}
  \includegraphics[width=7.8truecm]{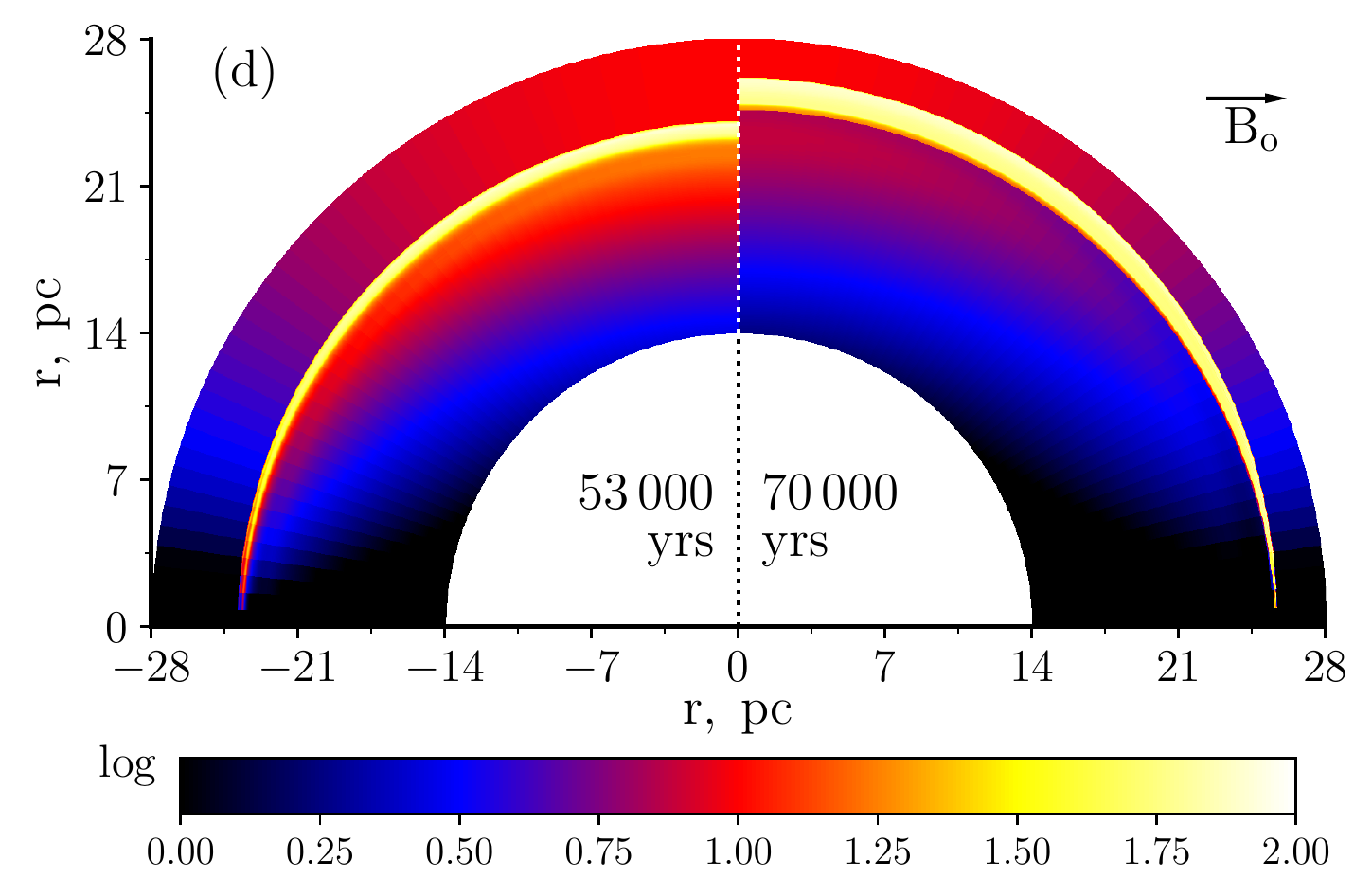}  
  \caption{The same as on Fig.~\ref{rad2:fig_angular1_oblique} for 53 000 yrs and 70 000 yrs. White line on the plot ({\bf c}) marks the shock location.
  }
  \label{rad2:fig_angular2_oblique}
\end{figure*}

The radial profiles of MHD parameters in the post-shock region are shown on Fig.~\ref{rad2:fig_profiles_oblique}.  Again, the distributions for the larger obliquities are similar to distributions downstream of the perpendicular shock with the larger MF strength (cf. Figs. 9, 10, 12 in Paper I). The tangential MF component compressed on the shock creates a sort of a pressure buffer in place of the decreasing (due to the radiative losses) thermal pressure. 
Since the tangential magnetic field component increases with the angle $\Theta\rs{o}$, the effect of the `pressure buffer' is larger for larger values of $\Theta\rs{o}$.
This effect obviously diminishes for shocks with small angles $\Theta\rs{o}$ (red lines on Fig.~\ref{rad2:fig_profiles_oblique}) and small strengths $B\rs{o\perp}$ (red lines on Figs.~9 and 10 in Paper I), allowing the highly dense radiative shells formation. The radial MF component rapidly decreases downstream of the radiative shock (bottom row plots on Fig.~\ref{rad2:fig_profiles_oblique}): the smaller $\Theta\rs{o}$ the thinner the radiative shell is and therefore the sharper the post-shock drop of $B\rs{\|}$ (Sect.3.3 in Paper I). The strengths of the MF components in the shell behind the shock may be estimated as $B\rs{\perp}\sim B\rs{o}\sin\Theta\rs{o}\big(R/R\rs{tr}\big)\big(\rho\rs{max}/\rho\rs{o}\big)$ and $B\rs{\|}\sim B\rs{o}\cos\Theta\rs{o}\big(R\rs{tr}/R\big)^2$ where $R\rs{tr}$ and $R$ are the shock radii at the time $t\rs{tr}$ and at the present time ($R\rs{tr}\approx 22.4 \un{pc}$ for the parameters considered).

The effects of the radiative losses and the shock obliquity in development of structures downstream of the post-adiabatic shock are visible on the Fig.~\ref{rad2:fig_oblique_space_time_T} (temperature) and Fig.~\ref{rad2:fig_oblique_space_time_rho} (density). It is clear that the losses start to impact the flow at the same time independently on the angle between the ambient MF and the shock normal. It is also interesting to note that the first element which cools down first is at the same distance behind the shock, for any obliquity. 
After this first element, the neighboring areas cool as well, once the temperature lowers to $\sim 10^5\un{K}$ (that corresponds to the maximum in the cooling curve $\Lambda(T)$, Fig.~\ref{rad2:fig_cooling}). The reverse shock (visible behind cooled down regions on Fig.~\ref{rad2:fig_oblique_space_time_T} after 60 000 yrs) travels somehow faster and deeper downstream of the quasi-perpendicular shock because of higher MF pressure in the post-shock shell which pushes the reverse shock inwards.
One can clearly see the cold region behind the forward shock is wider for the more oblique shocks (Fig.~\ref{rad2:fig_oblique_space_time_T}). At the same time, the cold shell has larger density for the quasi-parallel than for the quasi-perpendicular shocks (Fig.~\ref{rad2:fig_oblique_space_time_rho}). Evidently, the presence of the tangential MF itself is responsible for those effects.

What is the detailed picture of the obliquity variation of MHD parameters? We have created a `reconstruction' of the 2-D cross-sections on Figs.~\ref{rad2:fig_angular1_oblique}-\ref{rad2:fig_angular2_oblique}. In order to produce these plots, we used a set of 1-D simulations for different $\Theta\rs{o}$ values and combined them; each 1-D profile is shown at the angle corresponding to its obliquity. It is clear that MF affects the flow when and where the plasma $\beta$ (defined as the ratio between the magnetic and thermal energy densities) grows to unity and above (Figs.~\ref{rad2:fig_angular1_oblique}b and \ref{rad2:fig_angular2_oblique}b). The plasma structures do not feel MF up to the late adiabatic phase (Figs.~\ref{rad2:fig_angular1_oblique}) because $\beta$ is rather small. Losses start to grow in the post-adiabatic phase causing MF compression and, thus, a rise in $\beta$ value (Fig.~\ref{rad2:fig_angular2_oblique}). The compression depends on the MF orientation, therefore the flow profiles smoothly vary with direction.

\section{Non-uniform ambient density}
\label{rad2:sect4}

In the previous section, we considered properties of the post-adiabatic MHD shocks in the uniform medium. 
How do those properties change in the ISM with a non-uniform density distribution? In this section the uniform MF is assumed, both in the direction and the magnitude. Also we consider only the boundary cases of the parallel and perpendicular shocks which, as it is shown in the previous section, are extendable to any intermediate obliquities.

For the non-uniform ambient density we choose the exponential distribution
\begin{equation}
 n\rs{o}(r) = n\rs{o}(0)\exp\left({r/h}\right)
 \label{rad2:densexp}
\end{equation} 
where $h$ is the length-scale of the distribution (the density grows more rapidly for smaller $h$). We take into account only the evolution in the increasing density gradient medium because we are interested in SNR interactions with the increasing density medium such as molecular clouds. Such conditions facilitate shock deceleration which result in more prominent radiative losses.

The pressure in the ambient medium is related to the density and temperature by ideal gas law, $P\rs{o}\propto n\rs{o}T\rs{o}$. Since, the ISM density is not uniform, we performed simulations for two opposite scenarii: either $T\rs{o}=\mathrm{const}$ or $P\rs{o}=\mathrm{const}$. It appears that all parameter distributions downstream are almost the same in both cases, as expected, because in our simulations the ambient pressure is much smaller compared to the post-shock pressure, even for the strongest growth of $n\rs{o}$ and late integration times. The figures in this section are demonstrated for simulations with the constant ISM temperature. 

\begin{figure}
  \centering 
  \includegraphics[width=8.6truecm]{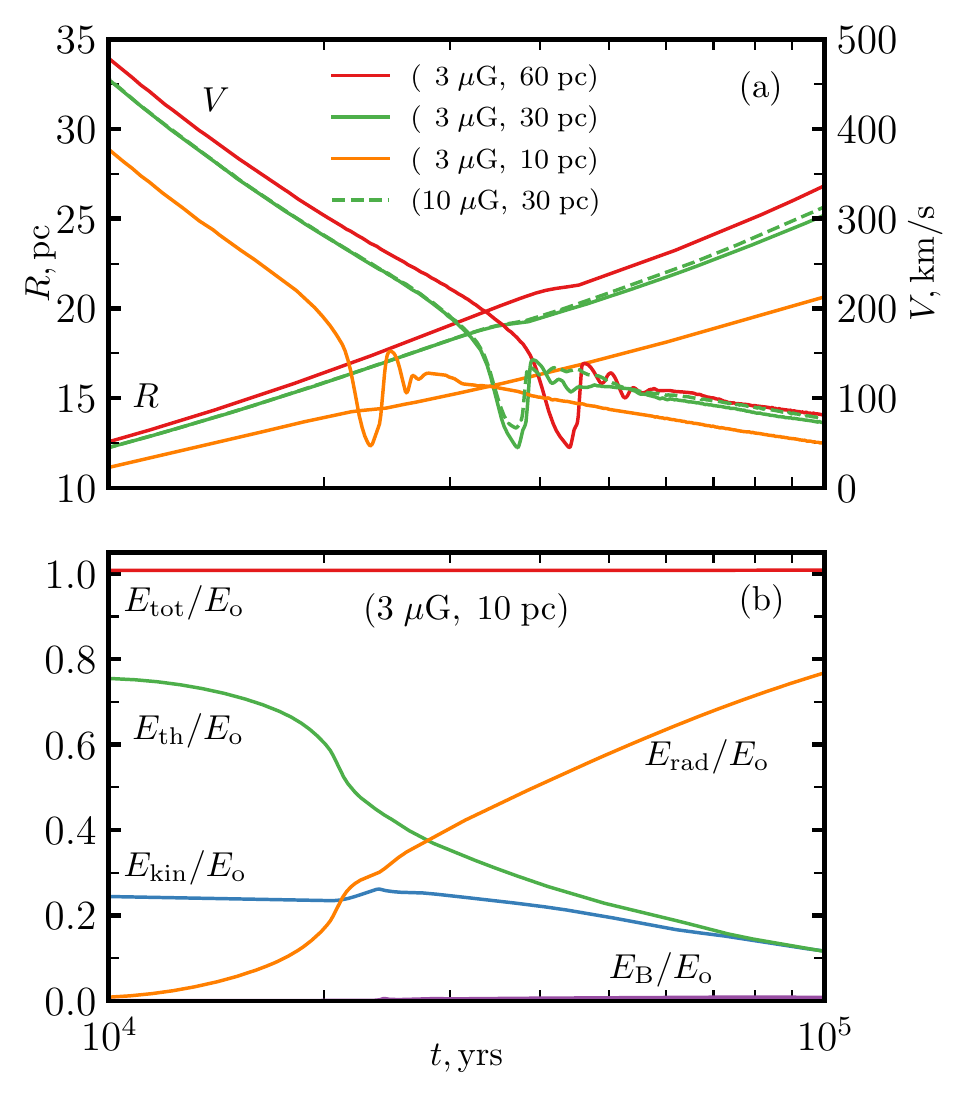}
  \caption{Effect of the ambient density gradient on the radius $R$ and velocity $V$ ({\bf a}) 
  of the perpendicular shock 
    for different MF strength $B\rs{o}$ and density scale $h$ values
    (given in parentheses). 
    Temporal evolution of the total energy components ({\bf b}) is shown for the model with 
    $h = 10\un{pc}$ and $B\rs{o} = 3\un{\mu G}$. 
  }
  \label{rad2:fig_grad_rho_blast_wave}        
\end{figure}

Fig.~\ref{rad2:fig_grad_rho_blast_wave} demonstrates the perpendicular shock dynamics (cf. Fig.~4 in Paper I where the uniform ISM was considered). During the adiabatic phase, the increasing ambient density slows down the shock in an agreement with $V\propto \big(n\rs{o}(R)R^3\big)^{-1/2}$ \citep{1999A&A...344..295H}. So the shock temperature $T\rs{s}\propto V^2$ decreases to $\sim 10^5\un{K}$ sooner for smaller values of $h$. As a result, we see that the part of the SNR shock encountering the molecular cloud enters the post-adiabatic stage earlier than regions moving in the uniform medium. It looks like the increasing density medium acts as a temporal `scale' of radiative processes (lines of different colors on Fig.~\ref{rad2:fig_grad_rho_blast_wave}a) while the MF strength affects the dynamical features of processes (the solid and dashed green lines on the same figure) without affecting the time-scale.  

\begin{figure*}
  \centering 
  \includegraphics[width=14truecm]{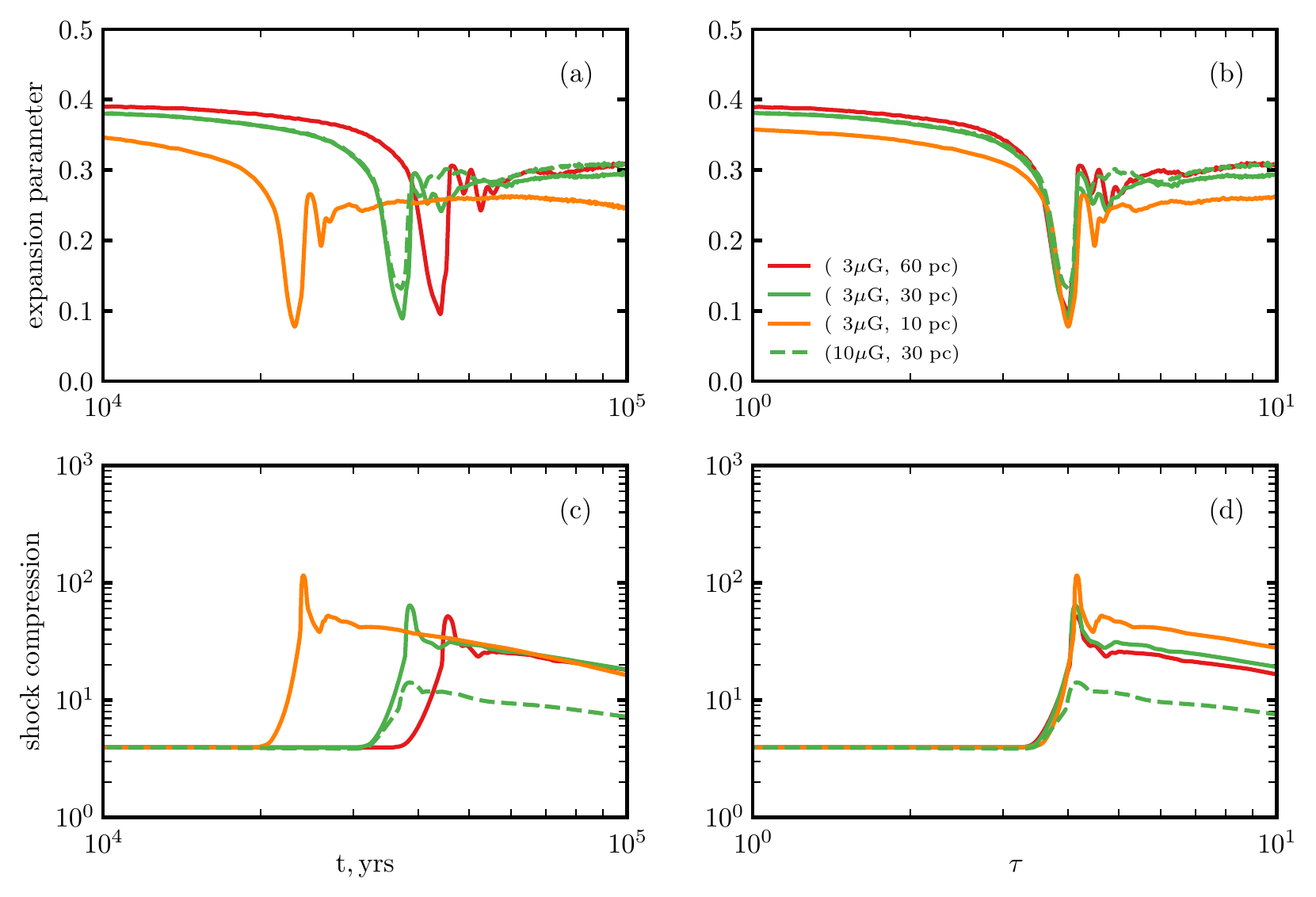}
  \caption{Temporal evolution of the expansion parameter $m$ ({\bf a}) and
    the shock compression factor $\rho\rs{max}/\rho\rs{o}$ ({\bf c}) for the {\it perpendicular} shock 
    in a medium (\ref{rad2:densexp}) with different density scales $h$. 
    Legend format: (MF strength $B\rs{o}$, length-scale $h$).
    Plots ({\bf b}) and ({\bf d}) correspond to ({\bf a}) and ({\bf c}) respectively, 
    with $\tau = 4t / t\rs{min}$ on the horizontal axis instead of $t$.
  }
  \label{rad2:fig_m_rho_max_grad_rho}
\end{figure*}
\begin{figure*}
  \centering 
  \includegraphics[width=14truecm]{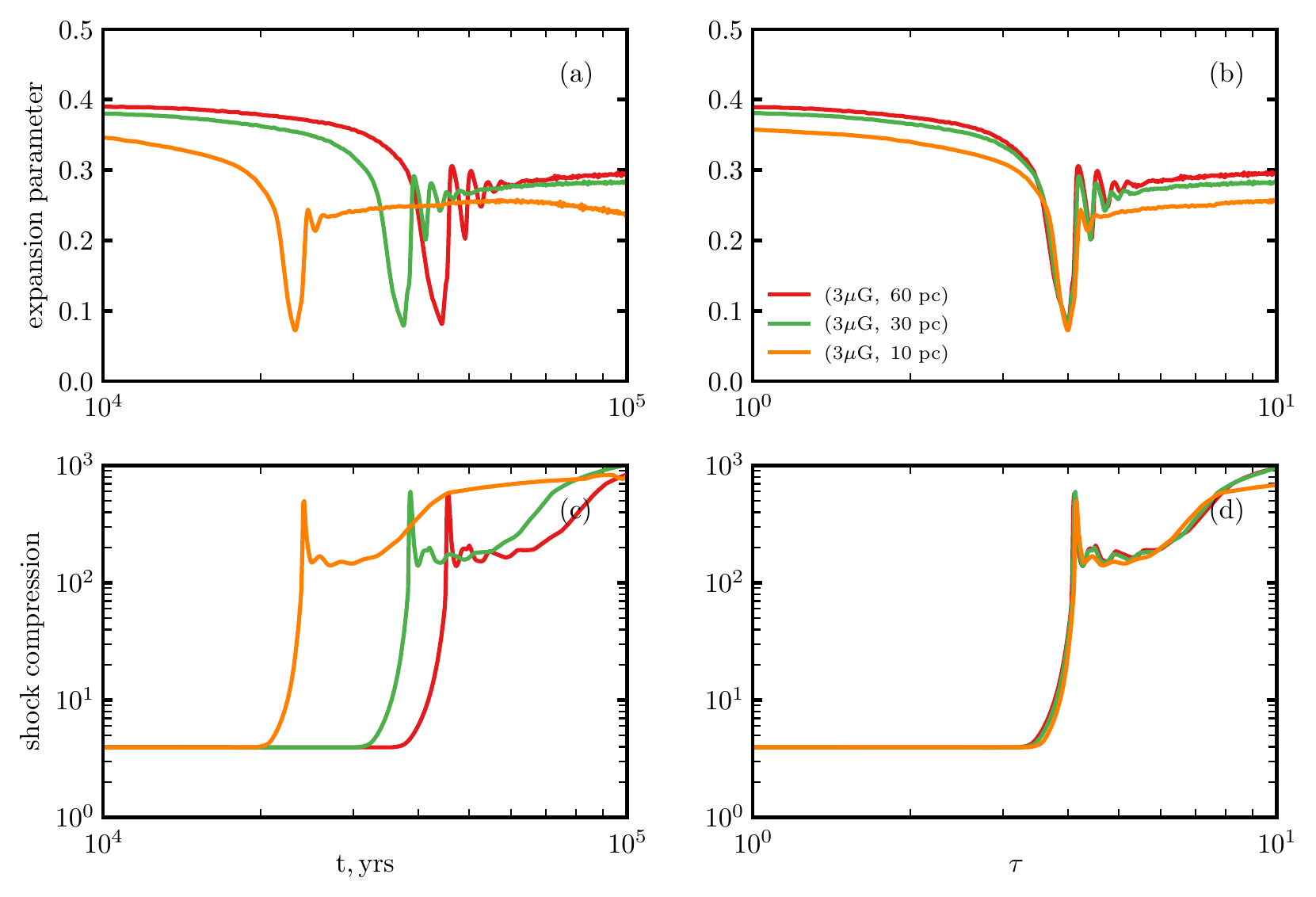}
  \caption{The same as on Fig.~\ref{rad2:fig_m_rho_max_grad_rho} for the {\it parallel} shock.
    The green dashed line is not displayed here because it should coincide with the green solid line 
    (MF does of affect the flow downstream of the parallel shock).
  }
  \label{rad2:fig_m_rho_max_grad_rho_parallel}
\end{figure*}

The energy transformations in the radiative MHD flows propagating in a non-uniform ISM (Fig.~\ref{rad2:fig_grad_rho_blast_wave}b for $h=10\un{pc}$) are qualitatively similar to the uniform medium case (Fig.~4c and d in Paper I, i.e. for $h\rightarrow \infty$) and in the ISM with other values of $h$. However, the transformations of the energy components due to the deviation from the adiabatic evolution begin at different times: later for larger $h$, in a full agreement with variations of $m$ (Fig.~\ref{rad2:fig_m_rho_max_grad_rho}a). This parameter is the best indicator for the evolutionary stage also for the shocks in a non-uniform medium. 

Indeed, the intersection of the yellow and blue lines on Fig.~\ref{rad2:fig_grad_rho_blast_wave}b  (energy radiated becomes compatible with flow kinetic energy) happens around time $t\rs{min}$ which marks the minimum value of $m$; the intersection of the yellow and green lines (a half of the thermal energy is radiated away) is right after the shell-formation time $t\rs{sf}$. Our simulations show that such a property holds approximately for a large range of values of the scale factor $h$. The processes indeed can be {\it re-scaled}. If we consider - instead of the real time $t$ - the reduced time $\tau$ defined as $\tau=4t/t\rs{min}$ (the factor $4$ has been chosen arbitrarily, for plotting purpose only), then shocks propagating in different ISM density gradients evolve in the same time-line (Fig.~\ref{rad2:fig_m_rho_max_grad_rho}). This property holds also for the parallel shock (Fig.~\ref{rad2:fig_m_rho_max_grad_rho_parallel}) and therefore for oblique shocks. 

The reason for the shorter radiative time-scale in a medium with a stronger density gradient is clear. The radiative losses become dynamically important around the `transition' time $t\rs{tr}$ which reduces for increasing densities because the radiative losses $L$ are proportional to $n^2$, Eq.~(\ref{rad2:radlossdef}). In addition, the shock decelerates more efficiently in higher densities, $V\propto n\rs{o}^{-1/2}$. Then the temperature $T\rs{s}\propto V^2$ falls faster as well causing the cooling coefficient $\Lambda(T)$ to approach the maximum (Fig.~\ref{rad2:fig_cooling}) on a shorter time-scale.

There is another property of the shock propagating in the non-uniform density visible on Fig.~\ref{rad2:fig_m_rho_max_grad_rho} and \ref{rad2:fig_m_rho_max_grad_rho_parallel}. Namely, the value of $m$ on the {\it adiabatic} stage (i.e. around $\tau=1$ on these plots) is not exactly $2/5$, as it should be for the Sedov shock in the uniform medium, but a little smaller. This property is known. In fact, the more general solution, for the point-like explosion in a medium with the power-law density $n\rs{o}(R)\propto R^w$, predicts the smaller adiabatic $m=2/(5+w)$ for $w>0$ \citep{1991ppt..book.....K}. As to the medium with exponential density distribution (\ref{rad2:densexp}), we may also derive an analytical expression for $m(R)$. Namely, with the use of the formula (3) for the shock velocity from \citet{1999A&A...344..295H}, we obtain: 
\begin{equation}
 m(R)=6\sqrt{2}(R/h)^{-5/2}F\left(\sqrt{R/2h}\right)+2(R/h)^{-2}(R/h-3)
 \label{rad2:mexpappr}
\end{equation}
where $F(x)$ is the Dawson function. 
Since $F(x)$ is expressed in terms of the special function, the more suitable is its simple approximation 
\begin{equation}
 m(R)\approx 0.35+0.05(1-R/h)
\end{equation}
which is the first order expansion of the expression (\ref{rad2:mexpappr}). It is quite accurate for $R<3h$; the maximum error is $10\%$ at $R=3h$. Thus, the deceleration parameter is smaller than $0.4$ during the adiabatic phase also for the exponentially increasing density. 

Once the radiative losses become important, $m$ decreases further, up to the time $t\rs{min}$ (Figs.~\ref{rad2:fig_m_rho_max_grad_rho} and \ref{rad2:fig_m_rho_max_grad_rho_parallel}) and continues to evolve in a manner similar to the uniform density case.

\begin{figure} 
  \centering 
  \includegraphics[width=7.8truecm]{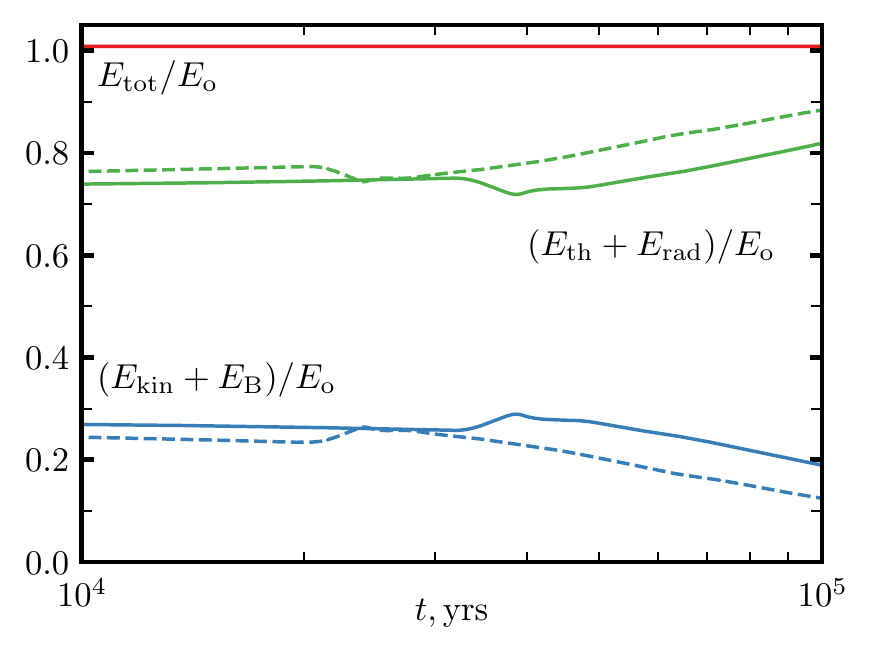}
  \includegraphics[width=7.8truecm]{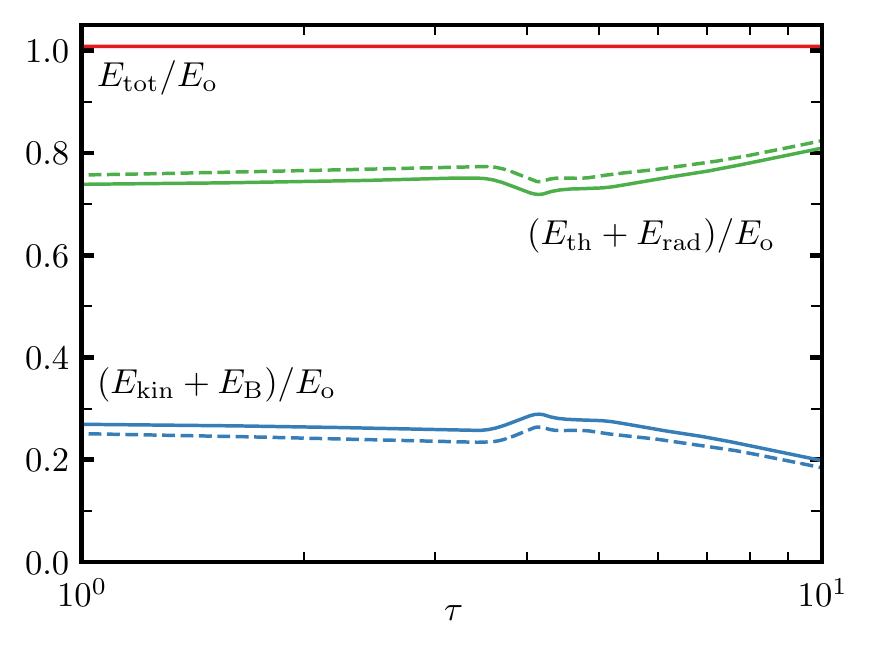}
  \caption{Temporal evolution of energy components sums.  
  The solid lines correspond to the model with the density lenght-scale $h = 30\un{pc}$ and the dashed lines represent $h = 10\un{pc}$. 
  $B\rs{o} = 3\un{\mu G}$. Perpendicular shock. 
  Lower plot is the same as the upper one but versus $\tau$ instead of $t$.
  }
\label{rad2:fig_grad_rho_energy_components}
\end{figure}

The evolution of the density compression $\rho\rs{max}/\rho\rs{o}$ by perpendicular and parallel shocks is shown on  Fig.~\ref{rad2:fig_m_rho_max_grad_rho}c and Fig.~\ref{rad2:fig_m_rho_max_grad_rho_parallel}c respectively. Again, the dependencies for different $h$ may be temporally re-scaled in order to see that the {\it physical} evolution is similar if considered in terms of $\tau$. After $t\rs{min}$, the compression on the parallel shock continues to evolve almost the same way,  even during the radiative phase: lines of different colors on Fig.~\ref{rad2:fig_m_rho_max_grad_rho_parallel}d -- which correspond to different density gradients -- almost coincide. However, the ratio $\rho\rs{max}/\rho\rs{o}$  is not able to reach high values at the perpendicular shock (Fig.~\ref{rad2:fig_m_rho_max_grad_rho}c) because the presence of MF pressure acts against the shock compression. The stronger perpendicular MF strength the more the shell compression is reduced (solid green versus dashed green line on Fig.~\ref{rad2:fig_m_rho_max_grad_rho}). The role of the perpendicular MF component diminishes for the stronger gradients of $\rho\rs{o}$ (i.e. for smaller $h$, Fig.~\ref{rad2:fig_m_rho_max_grad_rho}d, red to yellow lines) the faster deceleration of the forward shock. There are two reasons for such a property. First, the internal layers feel this deceleration with delay and fall onto the radiative shell with higher rate. Second, the stronger deceleration, the closer is $T$ to the maximum in $\Lambda(T)$, which implies the higher radiative losses and as a result the larger the shock compression.

The conversion of energy components is also a subject to re-scale, i.e. it follows very similar trends, though at different times, Fig.~\ref{rad2:fig_grad_rho_energy_components}. Namely, the drop in the thermal component gives rise to the thermal radiation while the flow kinetic energy transforms into the magnetic energy. These paths of energy exchange are almost independent (lines on Fig.~\ref{rad2:fig_grad_rho_energy_components} are close to horizontal), up to the well developed radiative stage (around $\tau=10$). 

The rescaling property may be useful for the understanding of internal processes or some parameters estimates. However, it does not mean that all processes develop in exactly the same way, regardless the density gradient, but just that the chain of the processes are similar because they are regulated by a common feature, radiative losses, which become effective when the shock velocity reaches some value. 

\begin{figure*}
  \centering 
  \includegraphics[width=13truecm]{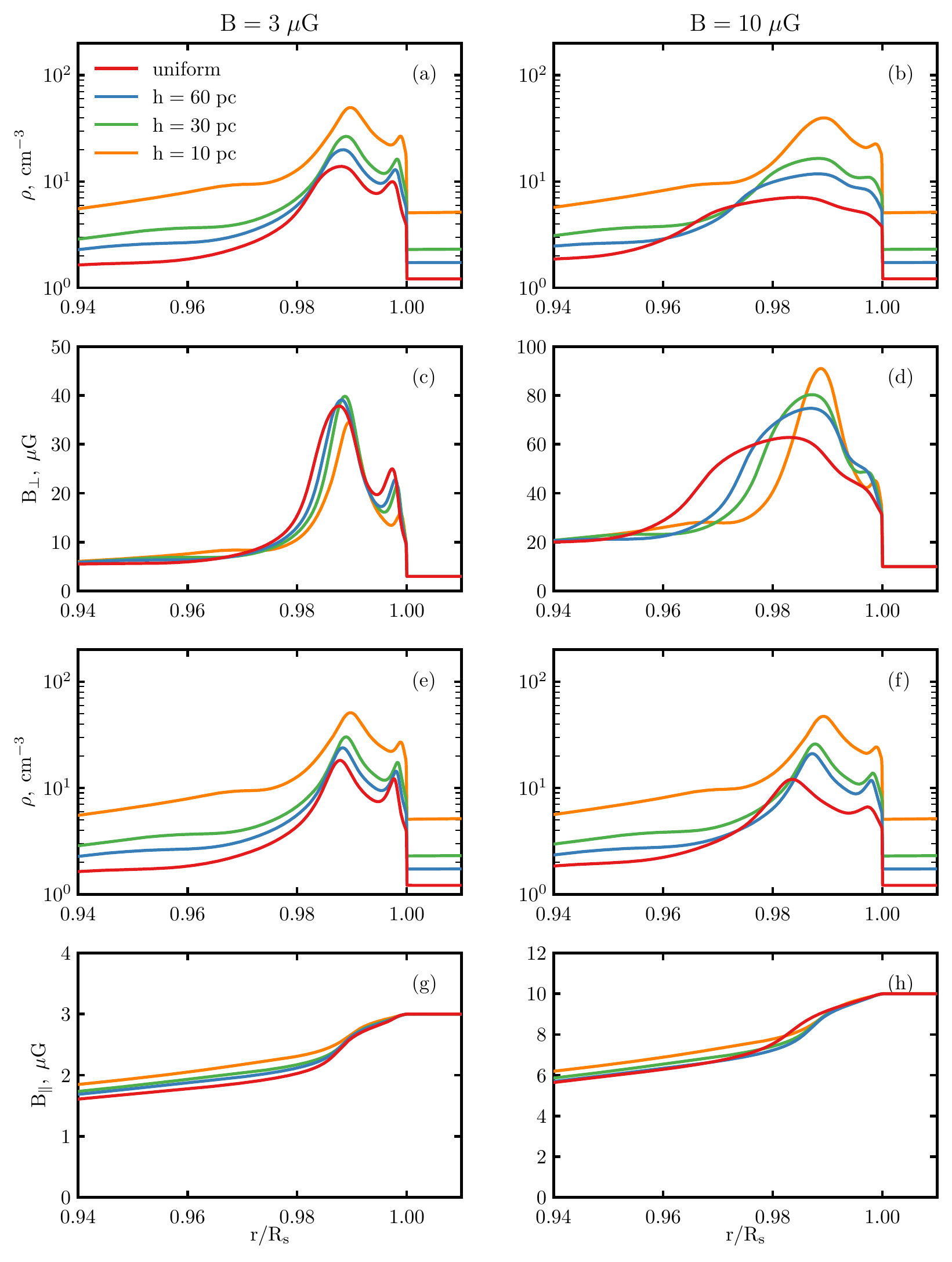}  
  \caption{Radial profiles of density and MF downstream of the perpendicular shock 
  ({\bf a}-{\bf d}) and behind the parallel one ({\bf e}-{\bf h}) for two values of the ambient MF 
  (left and right columns) and different scales of the density non-uniformity $h$.
  Profiles are shown for the different time moments which correspond however to the same evolutionary phase 
  at $t = 0.97 t\rs{min}$. 
  }
  \label{rad2:fig_grad_rho_profile}
\end{figure*}
\begin{figure*}
  \centering 
  \includegraphics[width=0.98\textwidth]{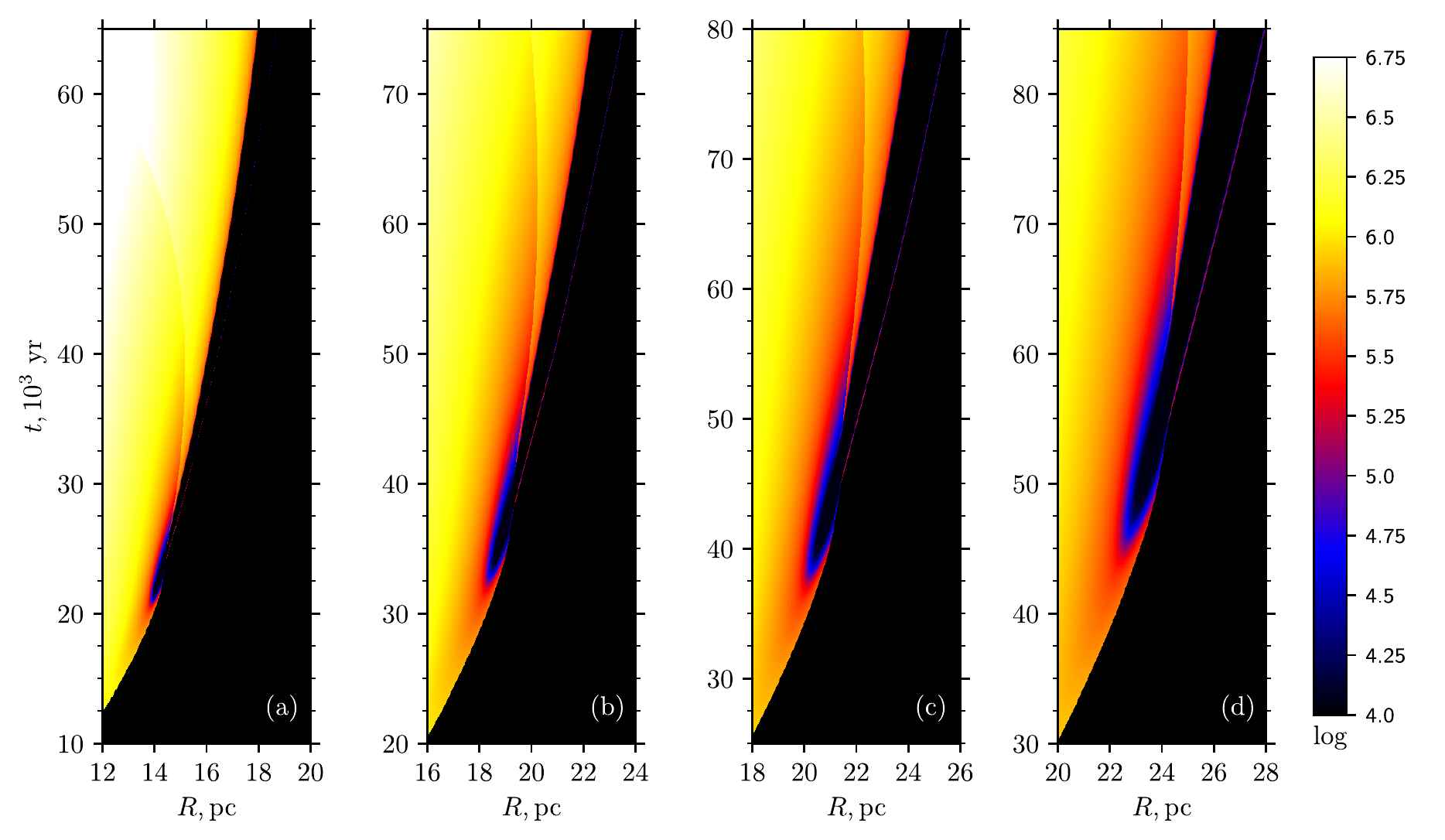}
  \caption{Time dependence of the temperature distribution behind the {\it perpendicular} shock 
    for the density scales $h = 10\un{pc}$ ({\bf a}), $h = 30\un{pc}$ ({\bf b}), $h = 60\un{pc}$ ({\bf c}), 
    and the model with the uniform ISM ({\bf d}). $B\rs{o} = 10\un{\mu G}$. 
    Plot d here represents the same parameter set as plot d on Fig.~6 in Paper I where the role of the MF 
    strength is demonstrated and 
    is the same as the right plot on Fig.~\ref{rad2:fig_oblique_space_time_T} displaying the role of 
    the shock obliquity. 
  }
  \label{rad2:fig_grad_rho_space_time_T}
\end{figure*}
\begin{figure*}
  \centering 
  \includegraphics[width=0.98\textwidth]{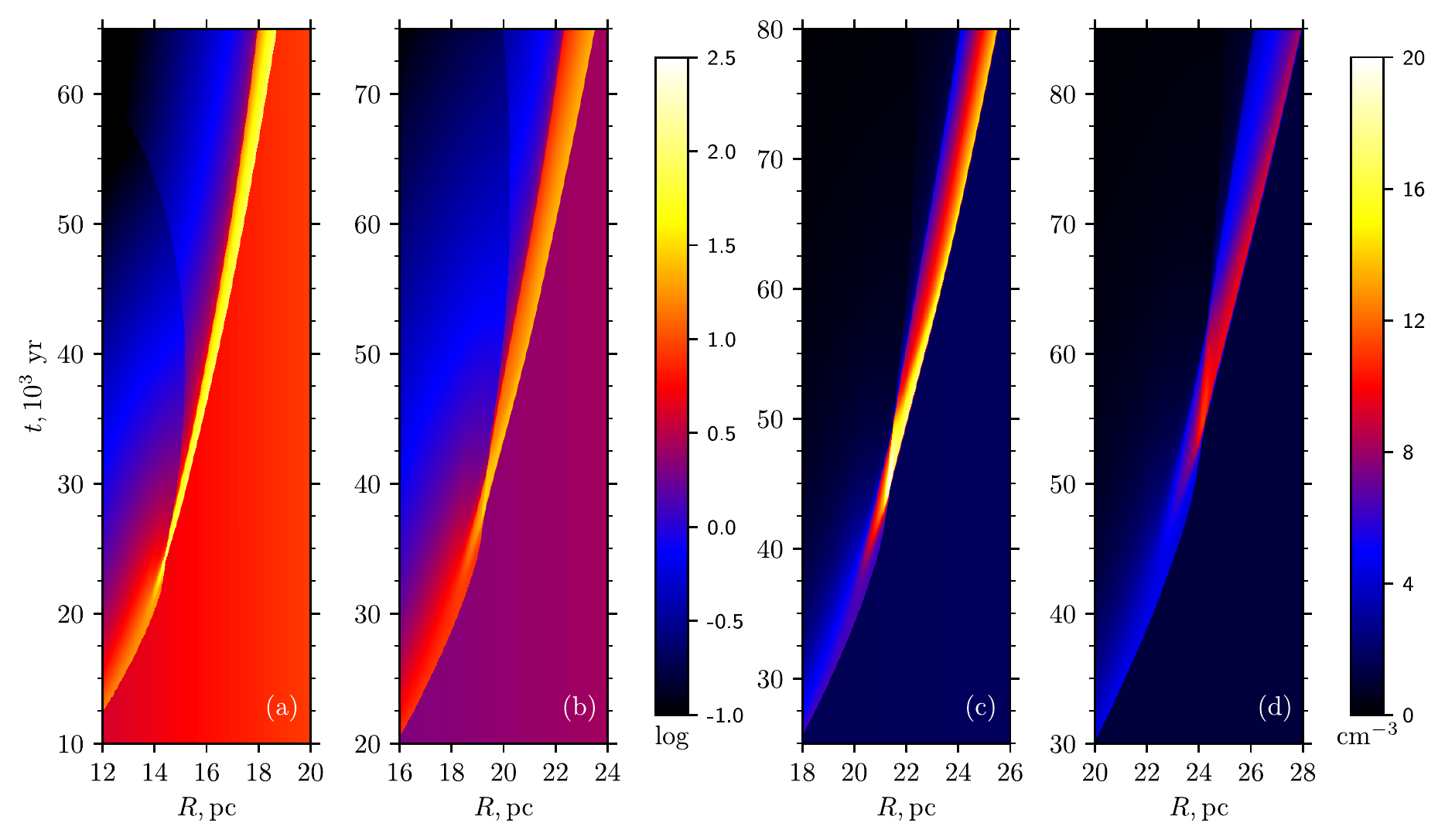}
  \caption{The same as Fig.~\ref{rad2:fig_grad_rho_space_time_T} for the number density  
  (cf. Fig.~7 in Paper I and Fig.~\ref{rad2:fig_oblique_space_time_rho} above). 
  }
  \label{rad2:fig_grad_rho_space_time_rho}
\end{figure*}

Some differences may be seen in the distributions of MHD parameters downstream of the perpendicular shock if MF strength is high enough to take prominent amount of energy from the shock, as it one can see on Fig.~\ref{rad2:fig_grad_rho_profile}b and d. (On Fig.~\ref{rad2:fig_grad_rho_profile}, we selected profiles 
at the same evolutionary phase). 
Instead, the structures in the radiative shell are quite similar for ISM with different density gradients if tangential MF is not so high (Fig.~\ref{rad2:fig_grad_rho_profile}a and c) or the parallel shock is considered (Fig.~\ref{rad2:fig_grad_rho_profile}e-h).  

\begin{figure}
  \centering 
  \includegraphics[width=\columnwidth]{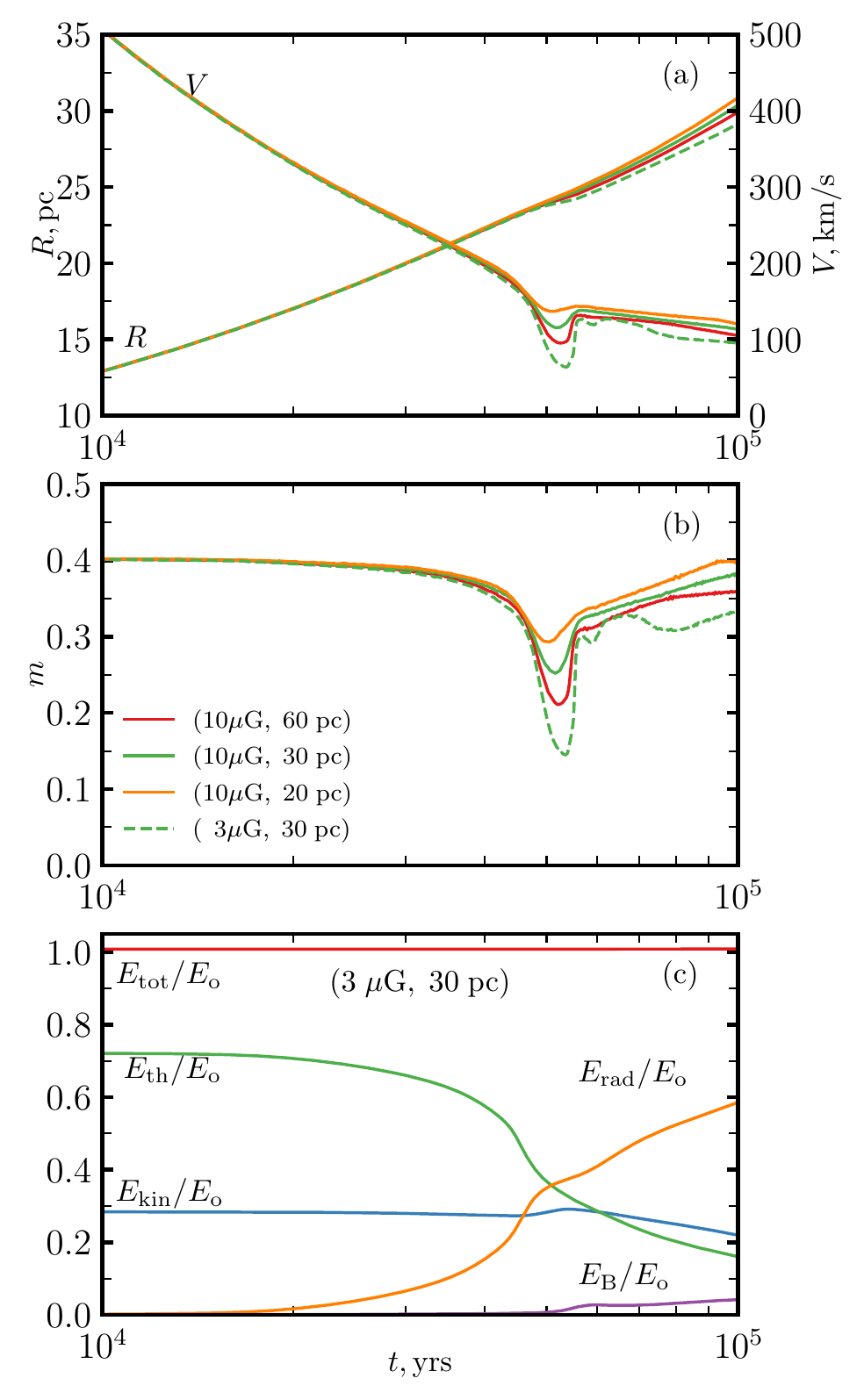}
  \caption{Time dependence of the shock and flow characteristics for 
  different strenght $B\rs{o}(0)$ and the length scale $H$ of the (tangential) ambient magnetic field. 
  $\rho\rs{o}(r)=\mathrm{const}$.
  }
  \label{rad2:fig_grad_B_blast_wave}        
\end{figure}

Note, that the horizontal axes on Fig.~\ref{rad2:fig_grad_rho_profile} are normalized to the shock radii $R(t)$ which are different for different lines, i.e. the thickness of shells (being similar on the plot) is different in absolute units. This is a sort of another scaling property, spatial. It is also interesting that density profiles on Fig.~\ref{rad2:fig_grad_rho_profile}a and e (i.e. if tangential MF is not large) may also be scaled on the vertical axis, by dividing by the value of $\rho\rs{o}(R)$. 

The evolution of structures, in particular the post-shock shell formation and reverse shock propagation, can be seen on Figs.~\ref{rad2:fig_grad_rho_space_time_T} and \ref{rad2:fig_grad_rho_space_time_rho} where 2D plots of temporal profiles dependence are shown in physical units. 
We see that similar features are developing behind the shock on all four plots, but on different time- and length-scales. 
For example, the blue feature outlines the region in the phase space where the radiative losses are effective. This region roughly outlines the duration of the post-adiabatic stage, which appears to be shorted for smaller values of $h$. The post-shell thickness on the radiative stage decreases with the decrease of $h$. At the same time the thickness of any specific radiative shell remains constant during its evolution. In some sense the presence of the density gradient results in suppressing the magnetic pressure effect on the shock (i.e. the shell becomes more thick and dense).

It looks like the right plot (for larger $h$) is a `close-up' version of the left plot (on the temporal axis). 
In fact, the leftmost Fig.~\ref{rad2:fig_grad_rho_space_time_T}a, due to the scaling property, presents the wider picture of the flow evolution. This explains in particular, why the reverse shock moves deeper downstream on this plot.

\section{Non-uniform ambient magnetic field}
\label{rad2:sect5}

In this section, we consider the SNR development in a medium with the uniform density distribution but with the MF strength increasing exponentially:
\begin{equation}
 B\rs{o}(r) = B\rs{o}(0)\exp\left({r / H}\right)
 \label{rad2:nismfdef}
\end{equation}
The only results for the perpendicular shock are presented in this section  because the tangential MF produces the strongest effect on the flow dynamics.

Based on Fig.~\ref{rad2:fig_grad_B_blast_wave}a one can conclude that
the shock radius and velocity are not strongly affected by the presence of
the MF gradient even with a relatively small scale ($H>20$ pc) and rather high $B\rs{o}(0)$. 
Some differences in the value of $V$ are present around $t\rs{min}$ only. 
Contrary to the evolution in the density gradient environment, the shock dynamics in the non-uniform MF has the same temporal milestones: minimum of the deceleration parameter occurs at the same time $t\rs{min}$ for different $H$  (Fig.~\ref{rad2:fig_grad_B_blast_wave}b). It happens because the cooling coefficient $\Lambda$ does not depend on $B$. Yet, minimum values of $m(t)$ around $t\rs{min}$ differ due to different values of ISMF strength $B\rs{o}(R)\propto \exp\left(R(t\rs{min})/H\right)$ before the shock front at this time (in the agreement with Fig.~\ref{rad2:fig_m_rho_max_oblique}a in the present paper and Fig.~4b in Paper I). 

Due to smaller shock deceleration (caused by the high post-shock MF pressure), the kinetic energy (Fig.~\ref{rad2:fig_grad_B_blast_wave}c)
is higher compared to the non-uniform ISM (Fig.~\ref{rad2:fig_grad_rho_blast_wave}b) cases even though a part of it is converted into MF energy.
The sums of the energy components of the system evolve in a similar manner as on Fig.~5 in Paper I, namely, $E\rs{kin}+E\rs{B}\approx\mathrm{const}$ and $E\rs{th}+E\rs{rad}\approx\mathrm{const}$. 
In other words, even if the ambient MF is non-uniform, the tangential MF still `drains' the energy from the kinetic component only while the radiative losses affect only the thermal energy. 

\begin{figure*}
  \centering 
  \includegraphics[width=0.98\textwidth]{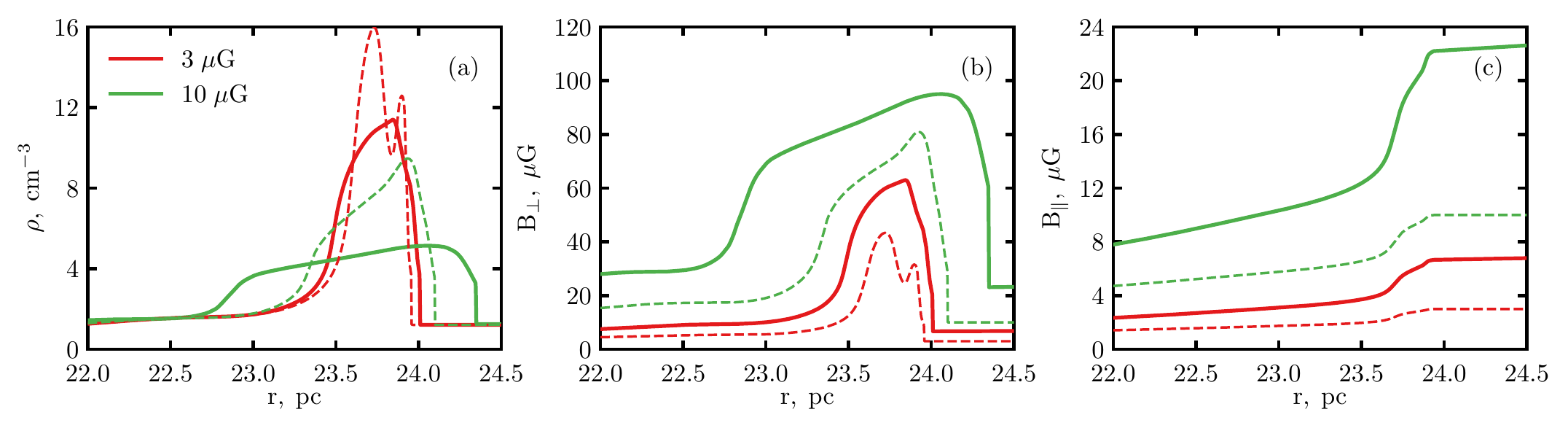}
  \caption{Radial profiles of density ({\bf a}) and MF ({\bf b}) downstream of the perpendicular shock 
  as well as MF behind the parallel shock ({\bf c}). 
  Solid lines are for the non-uniform MF (equation \ref{rad2:nismfdef} with $H=30\un{pc}$), 
  dashed lines represent analogous models in the uniform ISMF. 
  The strength $B\rs{o}(0)$ is shown on the plot legend, $\rho\rs{o}(r)=\mathrm{const}$, 
  $t = 53\;000$ yrs (that is $0.97t\rs{min}$)
  The density profiles for the parallel shock are similar to the red lines on ({\bf a}). 
          }
  \label{rad2:fig_grad_B_profile}
\end{figure*}

The shock compression $\rho\rs{max}/\rho\rs{o}$ 
does depend on the MF gradient scale $H$: at the same moment of time it is lower in models with lower values 
of $H$ (i.e. for more rapid growth of $B\rs{o}(R)$). 
This is a result of the increasing tangential MF pressure in the post-shock
region which prevents the shell compression while the effective radiative 
cooling takes place. 
These conclusions are clearly illustrated by the post-adiabatic
density and MF profiles (Fig. \ref{rad2:fig_grad_B_profile})
where exponential ISMF models are opposed to the constant ISMF models.

\begin{figure*}
  \centering 
  \includegraphics[width=0.98\textwidth]{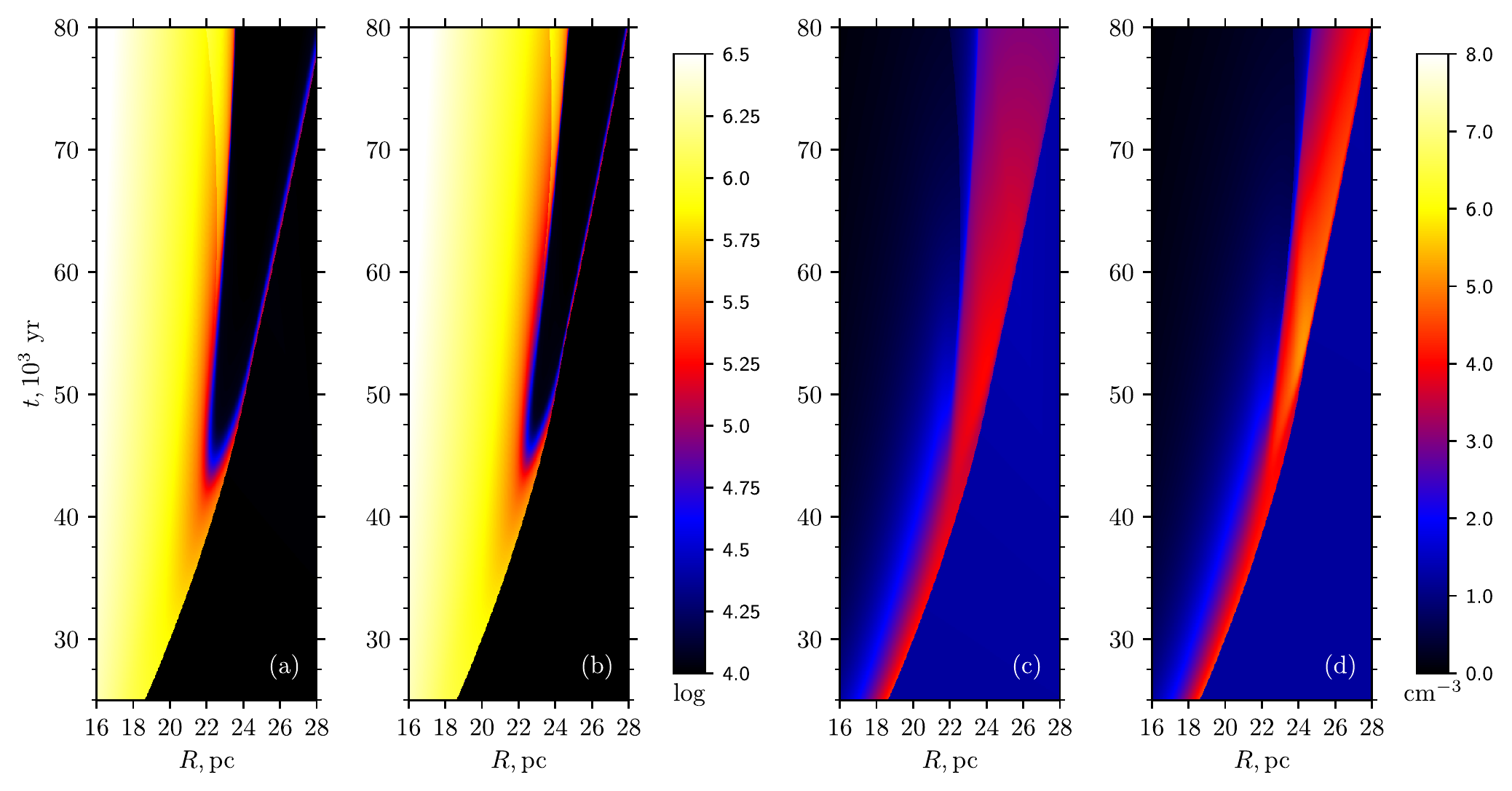}
  \caption{Time dependence of the distributions of temperature ({\bf a}, {\bf b})
    and density ({\bf c}, {\bf d}) behind the perpendicular shock in ISM with uniform density and 
    the exponential ISMF distribution with $H = 20\un{pc}$ ({\bf a}, {\bf c}) and 
    $H = 60\un{pc}$ ({\bf b}, {\bf d}).  
    $B\rs{o}(0) = 10\un{\mu G}$. 
  }
  \label{rad2:fig_grad_B_space_time}
\end{figure*}

Comparing Fig.~\ref{rad2:fig_grad_B_space_time} with corresponding plots for the uniform ambient MF  (Figs.~\ref{rad2:fig_oblique_space_time_T}d and \ref{rad2:fig_oblique_space_time_rho}d) we see that the radiative perpendicular shell in a nonuniform magnetic field is more `blurred': it is wider, less compressed and has less small-scale features. 
Also, because of the MF gradient the amount by which the shell is `blurred' increases with time 
(Fig.~\ref{rad2:fig_grad_B_space_time}b). It happens because of the rise of the post-shock MF strength, which is connected to the increasing pre-shock MF by the tangential MF jump conditions. Despite this, the amount of energy radiated is still high (Fig.~\ref{rad2:fig_grad_B_blast_wave}c), indicating that even high pre-shock MF has no impact on the thermal post-shock processes.

In summary, the presence of MF non-uniformity has significant
role in the post-shock shell structure modification but weakly affects the shock dynamics, even on the radiative phase.

\section{Non-Uniform ISM density with Non-Uniform MF}
\label{rad2:sect6}

Till now, we have considered the effects of the non-uniform density in ISM (Sect.~\ref{rad2:sect4}) and the non-uniform ambient magnetic field (Sect.~\ref{rad2:sect5}) independently. In the present section, we report simulations where the blast wave moves in the ISM where both $n\rs{o}(r)$ and $B\rs{o}(r)$ are not constant. Such a configuration is more likely in the situation when SNR interacts with a molecular cloud. We assume that density and MF strength follow equations (\ref{rad2:densexp}) and (\ref{rad2:nismfdef}) respectively. As to the respective length-scales, we take them to be related by the equilibrium condition between the thermal and magnetic pressure in the isothermal medium, which implies $H=2h$.

\begin{figure*}
  \centering 
  \includegraphics[width=15.6truecm]{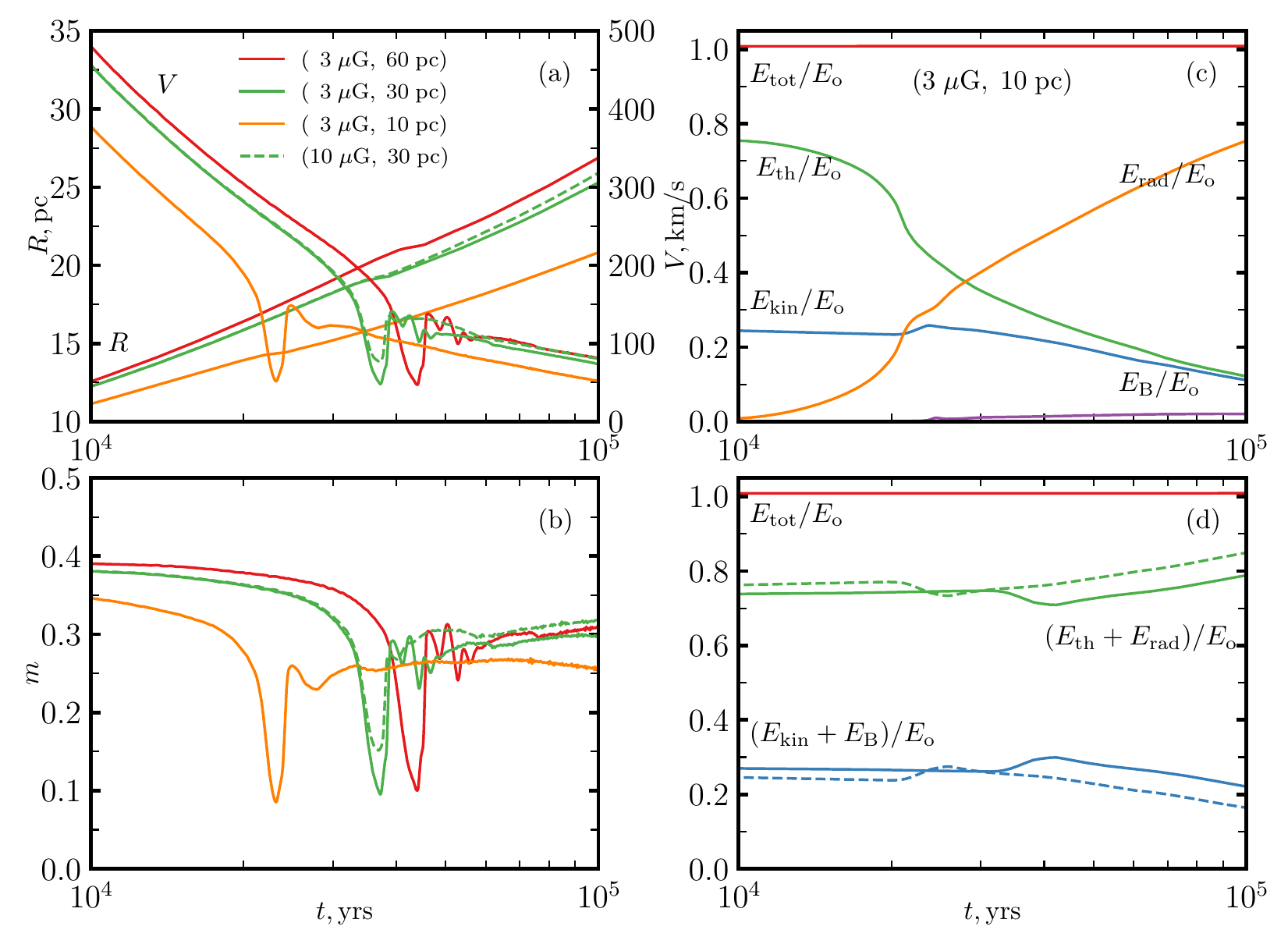}
  \caption{Time dependence of $R$, $V$ ({\bf a}), $m$ ({\bf b}), 
  energy components ({\bf c}) and their sums ({\bf d}). 
  Shock in medium with exponential distributions of density and MF, with $H=2h$.
  The pairs of parameters ($B\rs{o}(0)$, $h$) are marked on the plots a-c. 
  On the plot d, 
  the solid lines correspond to the model with ($3\un{\mu G}$, $30\un{pc}$) and 
  the dashed lines represent ($3\un{\mu G}$, $10\un{pc}$).
  }
  \label{rad2:fig_grad_rho_grad_B_blast_wave}        
\end{figure*}

The shock dynamics (Fig.~\ref{rad2:fig_grad_rho_grad_B_blast_wave}a, b) in such a model is very similar to the model described in Sect.~\ref{rad2:sect4} (Fig.~\ref{rad2:fig_grad_rho_blast_wave}, \ref{rad2:fig_m_rho_max_grad_rho}): the radius and velocity of the SNR's shock `feel' only the ISM density gradient. The minimum of $m(t)$ happens at the time which is determined by $\nabla \rho\rs{o}$. The shock is almost insensitive to the non-uniformity of the ambient MF, even if the strength of tangential MF is rather large. The (small) difference between models with $\nabla B\rs{o}\neq 0$ and $\nabla B\rs{o}=0$ consists in the evolution of the total magnetic energy (cf. Fig.~\ref{rad2:fig_grad_rho_grad_B_blast_wave} with Fig.~\ref{rad2:fig_grad_rho_blast_wave}, \ref{rad2:fig_m_rho_max_grad_rho} and \ref{rad2:fig_grad_rho_energy_components}): increasing by means of shock jump conditions while propagating in the increasing ISMF, as in model of Sect.~\ref{rad2:sect5} (cf. Fig.~\ref{rad2:fig_grad_B_blast_wave}). 

\begin{figure*}
  \centering 
  \includegraphics[width=0.98\textwidth]{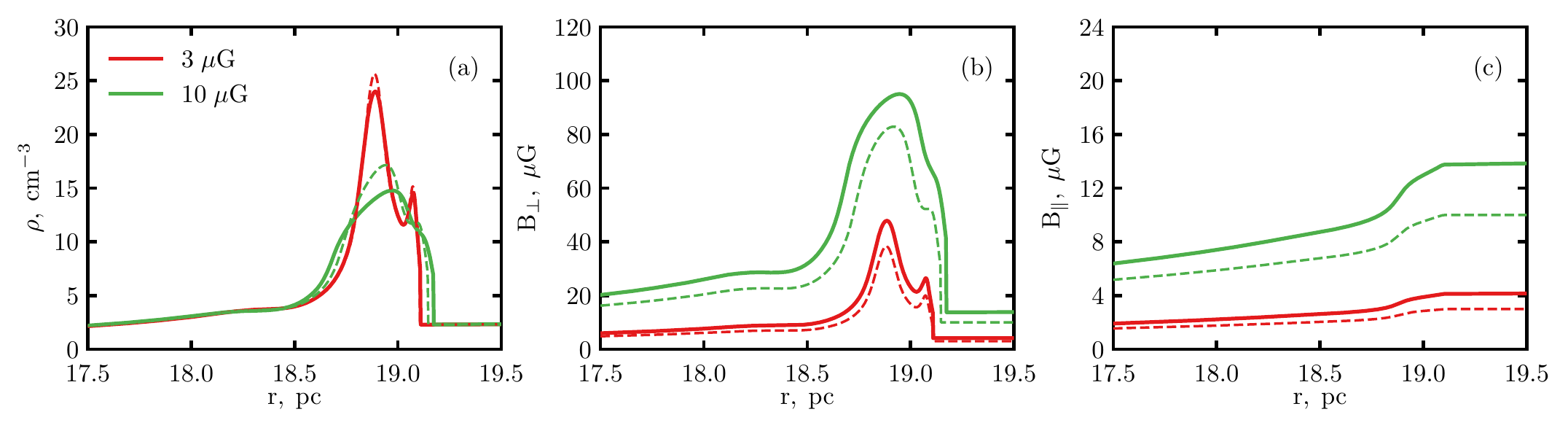}
  \caption{The same as Fig.~\ref{rad2:fig_grad_B_profile} for a shock wave in 
  non-uniform density with the scale $h=30\un{pc}$. 
  Dashed lines are for model with $B\rs{o}=\mathrm{const}$ 
  (the same as green lines on on Fig.~\ref{rad2:fig_grad_rho_profile}),
  solid lines correspond to model with non-uniform MF with $H=2h$. 
  Time $t = 36\;000$ yrs that corresponds to $0.97t\rs{min}$. 
  The range of $r$ on the horizontal axis here and on Fig.~\ref{rad2:fig_grad_B_profile} 
  shows about $10\%$ of the shock radius.
  }
  \label{rad2:fig_grad_rho_grad_B_profile}
\end{figure*}

The presence of the MF gradient contributes to the modification of the post-shock structures. 
Solid lines on Fig.~\ref{rad2:fig_grad_rho_grad_B_profile} demonstrate somehow wider radiative shell with smaller density peaks (behind perpendicular shock) and larger $B$ downstream compared to the model with $\nabla B\rs{o}=0$ (dashed lines). Comparing to Fig.~\ref{rad2:fig_grad_B_profile}, the thickness of the shell (in units of the shock radius) on Fig.~\ref{rad2:fig_grad_rho_grad_B_profile} is smaller because the MF gradient is not strong enough: the MF length-scale is $H=60\un{pc}$ on Fig.~\ref{rad2:fig_grad_rho_grad_B_profile} and $30\un{pc}$ on Fig.~\ref{rad2:fig_grad_B_profile}.

\begin{figure*}
  \centering 
  \includegraphics[width=0.98\textwidth]{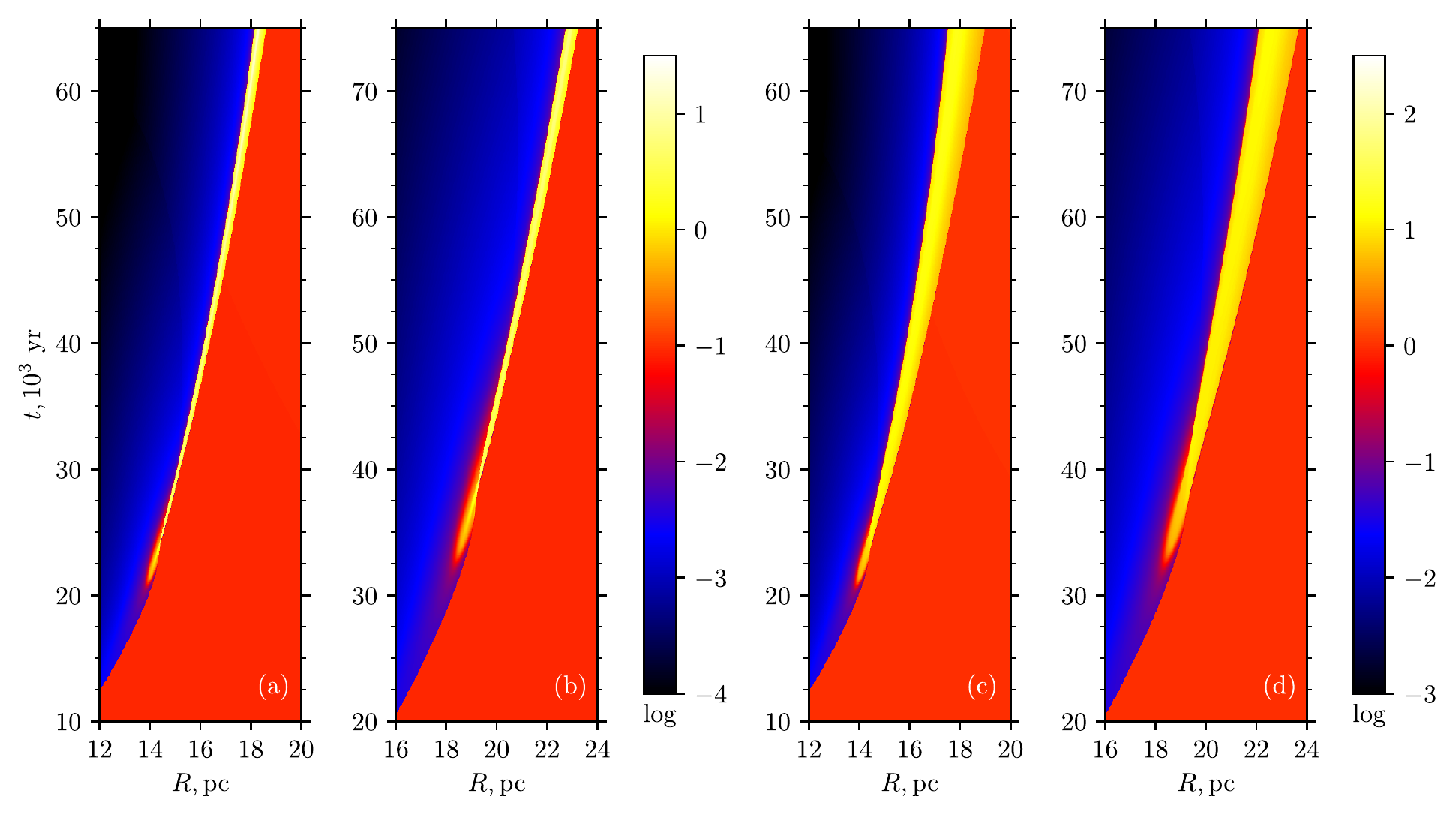}
  \caption{Evolution of the distributions of the plasma $\beta$ for the shock in nonuniform ISM with $H=2h$ 
  for models with the following parameters ($B\rs{o}(0)$, $h$): 
         ({\bf a}) ($3\un{\mu G}$, $10\un{pc}$), 
         ({\bf b}) ($3\un{\mu G}$, $30\un{pc}$), 
         ({\bf c}) ($10\un{\mu G}$, $10\un{pc}$), 
         ({\bf d}) ($10\un{\mu G}$, $30\un{pc}$). 
  }
  \label{rad2:fig_grad_rho_space_time_beta}
\end{figure*}

Magnetic field does not affect the evolution of SNRs up to the end of adiabatic phase because the thermal pressure $P$ is much higher than the magnetic pressure $P\rs{B}$. The MF becomes important only at the post-adiabatic phase when the radiative losses make plasma more compressible due to the thermal pressure drop. The compression of the tangential MF (frozen-in the plasma)  leads to the MF pressure increase up to the level $P\rs{B}>P$. This is demonstrated on Fig.~\ref{rad2:fig_grad_rho_space_time_beta} where the evolution of $\beta\equiv P\rs{B}/P$ is shown. 

\section{Conclusions}

Magnetic field at non-radiative SNR shocks has a small energy density compared to the kinetic or thermal energy component. Therefore, MF structure may be modelled independently of the distribution of the hydrodynamical parameters inside SNR, i.e. MF may be simulated with Maxwell equations on the hydrodynamic background.

With the end of the adiabatic phase, magnetic field becomes an important factor in the SNR dynamics. This happens because the radiative losses lead to the increase of the tangential MF due to the increased plasma compression in the radiative shell. 
Therefore, models which consider the interactions of SNRs with molecular clouds (or SNR evolution after the adiabatic stage in general) have to take magnetic field into account. 

We considered the role of MF in the evolution of SNR shocks which experience energy losses through thermal radiation. We also studied how the non-uniformities of density and MF modify the dynamics of the (partially and fully radiative) shock and structure of the flow. In particular,
\begin{itemize}
\item The dynamical influence of MF on the radiative shock originates from the tangential MF component that provides pressure support in the (partially) radiative shell. The flow structure behind an oblique post-adiabatic shock is similar to that downstream of a perpendicular shock, provided its tangential MF strength is the same as that in the oblique case.
\item The tangential MF prevents the large compression at a post-adiabatic shock which is present in the purely hydrodynamic simulations.
\item In situations covered by our model, i.e. when SNR interacts with the large-scale cloud, cosmic rays that are either newly accelerated or reaccelerated from the Galactic pool will not have so hard spectrum as in the simulations without magnetic field for two reasons. First, the compression ratio will not get very large and, second, the radiative shell is thicker, implying that only high-energy particles with large mean free-path can experience significantly enhanced compression. 
\item The large-scale non-uniformity of density and MF distributions in the vicinity of a molecular cloud modifies the evolution of an SNR as, e.g., an increasing ISM density speeds up the shock evolution meaning the beginning of the post-adiabatic regime may happen at younger age. 
\item One of the most interesting effects of a density gradient is an approximate rescaling of the temporal evolution and the spatial post-shock structures (Sect.~\ref{rad2:sect4}). 
\item The gradient of MF strength does not speed up the shock evolution, contrary to the ambient density gradient. 
\item The shock sweeps up more MF when moving in the direction of the positive gradient of the ambient MF. So, the effects of MF pressure in the radiative shell are more explicit, namely the shell becomes more `blurred' due to the increasing MF pressure which more efficiently prevents strong compression.
\end{itemize}

Magnetic field and its orientation are important also in the processes of the dense cloud formation. In 1D MHD simulations, oblique MF ($\Theta\gtrsim 30^\mathrm{o}$) prevents development of the thermal instabilities which are necessary for birth of dense molecular clouds  \citep{2000A&A...359.1124H}. 2D MHD simulations of converging flows (leading to formation of clouds) have demonstrated properties similar to those found in the present paper. Namely,  \citet{2008ApJ...687..303I,2009ApJ...704..161I} considering slower (but still supersonic) oblique radiative shocks have revealed the magnetic pressure support of the post-shock layer which leads, in particular, to the lower shock compression and somehow larger shock radii if there is a prominent tangential MF component presented in a setup. 
3D simulations \citep{2009ApJ...695..248H} of cloud formation confirm that tangential MF limits  the high compression of radiative gas, contrary to the HD models. Our results, though  derived for the faster shocks, are in agreement with findings in the referred studies: it is the magnetic pressure which rises downstream of the (partially) radiative oblique shock to oppose the ram pressure, not the thermal pressure as in the pure HD simulations.

It is worth to note here that, given our 1D approach, we described the interaction of the blast wave with a molecular cloud by assuming a smooth increasing density medium like at the outskirts of molecular clouds. This corresponds to assume that the cloud has a size larger than the size of the remnant (large-scale non-uniformity). In the case of the small-scale clouds (smaller comparable with a size of the remnant) and the sharp density contrast between the ISM and the cloud, the shock-cloud interaction is expected to trigger the development of hydrodynamic instabilities. After the blast wave starts to envelope the cloud, the instabilities developing at the cloud border contribute to the gradual erosion and fragmentation of the cloud (see, for instance \citealt{1994ApJ...420..213K}) and may lead to the formation of regions dominated by radiative cooling which, in turn, may trigger the development of thermal instabilities (e.g. \citealt{2008ApJ...678..274O}). This process, however, cannot be described by 1D simulations and requires a multi-dimensional approach. At the same time, the accurate description of radiative shocks as those explored here requires a very high spatial resolution which largely increases the computational cost of multi-D simulations.  On the other hand, we note that, in the general case of oblique shocks (as those investigated here), the magnetic field gradually envelopes the small-scale cloud during the shock-cloud interaction. This leads to a continuous increase of the magnetic pressure and field tension at the cloud border and, ultimately, to the suppression of hydrodynamic instabilities (e.g. \citealt{2008ApJ...678..274O}). Thus, the results of our simulations are valid for the remnants interacting with large molecular clouds without sharp edges or of ambient magnetic field strengths able to dump the hydrodynamic instabilities that would develop at the border of the clouds.

Models where the radiative shock impacts the small dense clumps have to account that the shock is not radiative immediately after it hits the clump. It takes time for the flow to modify the structure to the level when the shock compression exceeds the common value for the adiabatic shock ($4$ for the adiabatic index $5/3$) and provides conditions for efficient
adiabatic re-acceleration of the pre-existing cosmic rays. In addition, the tangential MF and the efficiently (re-)accelerated particles limit the shock compression considerably. The shock compression factor evolves in time in the radiative phase, depending on a shape of the cooling curve at the low temperatures (Figs.~\ref{rad2:fig_m_rho_max_oblique}b). This affects the slope of the cosmic ray spectrum which depends on the compression.

The present paper, as well as the Paper I, reveals the physical reason for an observational property. Namely, the middle-age and old SNRs demonstrate predominantly tangential orientations of MF vectors in the radio polarization maps \citep{1976AuJPh..29..435D,2015A&ARv..23....3D}. 
We see in our simulations the increase of the tangential MF component and drop of the parallel MF component behind the post-adiabatic shock that leads naturally to the prevalence of the tangential MF over the most of SNR shell. 

\section*{Acknowledgements} 

We acknowledge the CINECA Awards HP10CKMKX1,2016 and HP10CR7V42,2017 
for the availability of the high performance computing resources and support. 
A part of simulations were performed on the computational clusters in DESY, Zeuthen, and in Institute for Applied Problems in Mechanics and Mathematics, Lviv. 

\bibliographystyle{mnras}
\bibliography{rad2refs} 
  
\bsp	
\label{rad2:lastpage}
\end{document}